\documentclass[preprint, preprintnumbers ,amsmath,amssymb]{revtex4}
\usepackage{amssymb}
\usepackage{amsmath}
\usepackage{graphicx}
\usepackage{multirow}

\begin{document}



\title{New Perspectives on the Schr{\"o}dinger-Pauli Theory of Electrons:
Part I}



\author{Viraht Sahni}

\affiliation{Brooklyn College and The Graduate Center of the City
University of New York, New York, New York 10016.}


\date{\today}

\begin{abstract}
Schr{\"o}dinger-Pauli (SP) theory is a description of electrons in
the presence of a static electromagnetic field in which the
interaction of the magnetic field with both the orbital and spin
moments is explicitly considered. The theory is described from a new
perspective, viz. that of the individual electron via its equation
of motion or `Quantal Newtonian' first law. The law leads to new
physical and mathematical insights into the theory. The law is in
terms of `classical' fields whose sources are quantum mechanical
expectation values of Hermitian operators taken with respect to the
system wave function. The law states that each electron experiences
an external and an internal field, the sum of which vanish. The
external field is the sum of the electrostatic and a Lorentz field.
The internal field is a sum of fields: the electron-interaction,
differential density, kinetic, and internal magnetic fields. These
fields are respectively representative of a property of the system:
electron correlations due to the Pauli exclusion principle and
Coulomb repulsion, the electron density, kinetic effects, and the
physical current density. The energy components can be expressed in
integral virial form in terms of these fields. The law leads to the
further understanding that the Hamiltonian is an exactly known and
universal functional of the wave function. This allows for the
generalization of the SP equation, which then proves it to be
intrinsically self-consistent. A Quantal density functional (local
effective potential) theory of the SP system is developed. Further
generalizations of the present work to the temporal case, and
relativistic Dirac theory are proposed.
\end{abstract}


\maketitle



\section{Introduction}

The Schr{\"o}dinger-Pauli theory \cite{1} is a description of a
quantum-mechanical system comprised of N electrons in the presence
of an external electrostatic field ${\boldsymbol{\cal{E}}}
({\bf{r}}) = - {\boldsymbol{\nabla}} v ({\bf{r}})/e$ and a
magnetostatic field ${\boldsymbol{\cal{B}}} ({\bf{r}}) =
{\boldsymbol{\nabla}} \times A({\bf{r}})$, where $v ({\bf{r}})$ and
$A ({\bf{r}})$ are scalar electrostatic and vector magnetic
potentials, in which the interaction of the magnetic field with both
the orbital and spin angular momentum is explicitly considered.  The
purpose of this paper is to describe Schr{\"o}dinger-Pauli theory of
the many-electron system from a new perspective, one that leads to
further physical insights into the system, and thereby of our
quantum-mechanical understanding of Schr{\"o}dinger-Pauli theory.
The perspective is that of the \emph{individual} electron via its
stationary state equation of motion --  the `Quantal Newtonian'
first law.  As a consequence, it is proved that the Hamiltonian is
an \emph{exactly known} and \emph{universal} functional of the wave
function. This generalizes the Schr{\"o}dinger-Pauli equation. The
generalized form of the equation in turn exhibits its intrinsic
self-consistent nature.  The electronic system as described by the
Schr{\"o}dinger-Pauli equation is then mapped via quantal density
functional theory \cite{2,3,4} to one of noninteracting fermions
possessing the same basic variables of the electronic density $\rho
({\bf{r}})$ and physical current density ${\bf{j}} ({\bf{r}})$, and
from which the same total energy can be obtained. The mapping allows
for the determination of additional properties of the system not
obtainable solely by solution of the Schr{\"o}dinger-Pauli equation.
Hence, the noninteracting fermion model system constitutes an
essential complement to the Schr{\"o}dinger-Pauli theory.  The
noninteracting fermionic model is also an alternative description of
the physical system.  It is a local effective potential theory, and
as such it is more amenable to numerical solution.

The Schr{\"o}dinger-Pauli Hamiltonian is descriptive of a wide range
of physical phenomenon such as the Zeeman effect (weak, intermediate
and high magnetic fields); cyclotron resonance; magnetoresistance;
and the magneto-caloric effect \cite{5,6}.  For the two-dimensional
electron gas in semiconductor heterostructures, it is applicable to
the fractional quantum Hall effect \cite{7,8,9,10} at high magnetic
fields when the electrons become spin-polarized and the system
approaches an incompressible fluid. It is also applicable to the
harmonically bound two-dimensional `artificial atoms' or quantum
dots \cite{11,12,13,14} in such structures.  Such `artificial atoms'
are of particular interest as the modification of the energy
spectrum is discernable at magnetic fields of a few Tesla.  These
`artificial atoms' have electronic structure similar to that of
natural atoms \cite{11,12,13,14,15,16,17,18,19}. However, as the
size of the quantum dots is an order of magnitude greater than that
of natural atoms, the corresponding electronic density is low. As
such the electron correlations due to the Pauli exclusion principle
and Coulomb repulsion play a more significant role relative to the
kinetic energy than in natural atoms. Additionally, the contribution
of these correlations to the kinetic energy -- the
correlation-kinetic  energy -- becomes significant.  The confinement
of the electrons in `artificial atoms', and therefore the structure
of the electronic density, may also be altered experimentally so as
to allow for a study \cite{20,21,22,23,24} of the Wigner
\cite{25,26} high electron correlation regime of the two-dimensional
electronic system.  Wigner systems are characterized not only by a
high value of the electron-interaction energy relative to the
kinetic energy, but also by a high correlation-kinetic energy
\cite{22,23,24}. Wigner crystals in such systems can also be created
by strong magnetic fields \cite{27}: the electron correlations
become significant because the single particle states become
degenerate to form Landau levels with the electrons occupying the
lowest level. (Three-dimensional spherical quantum dots have also
been experimentally obtained \cite{28,29} and studied \cite{22,23}
in the Wigner regime.)  There has also been recent interest, both
experimental and theoretical, in studying yrast states for
harmonically bound electrons in a magnetic field \cite{14}. These
are states of lowest energy for fixed angular momentum.

The new perspective of the quantum system is that of the
\emph{individual} electron, in the sea of electrons, as described by
its equation of motion, the `Quantal Newtonian' first law for the
electron.  The law is in terms of `classical' fields that arise from
quantal sources.  The statement of the law is that the sum of the
external ${\boldsymbol{\cal{F}}}^\mathrm{ext} ({\bf{r}})$ and
internal ${\boldsymbol{\cal{F}}}^\mathrm{int} ({\bf{r}})$ fields
experienced by \emph{each} electron vanishes.  The external field
${\boldsymbol{\cal{F}}}^\mathrm{ext} ({\bf{r}})$ is a sum of the
external binding electrostatic ${\boldsymbol{\cal{E}}} ({\bf{r}})$
and the Lorentz ${\boldsymbol{\cal{L}}} ({\bf{r}})$ fields.  The
Lorentz field ${\boldsymbol{\cal{L}}} ({\bf{r}})$ depends upon the
cross-product of the physical current density ${\bf{j}} ({\bf{r}})$
and the magnetic field ${\boldsymbol{\cal{B}}} ({\bf{r}})$. (In
classical physics, the Lagrangian for a particle of charge $q$ in
the potentials $\{ v, {\bf{A}} \}$ contains the Lorentz force
explicitly \cite{2}.  However, in the corresponding Hamiltonian,
this term does not appear as it cancels out.  Hence, it does not
appear in the quantum-mechanical Hamiltonian obtained via the
correspondence principle.  Thus, in quantum mechanics, it is
implicitly understood that electrons in the presence of  a magnetic
field ${\boldsymbol{\cal{B}}} ({\bf{r}})$ experience a Lorentz
force.  In the `Quantal Newtonian' first law, the Lorentz field,
derived from the Lorentz `force', appears \emph{explicitly}.) The
internal field ${\boldsymbol{\cal{F}}}^\mathrm{int} ({\bf{r}})$ is a
sum of fields, each representative of a property of the system:
properties such as the correlations due to the Pauli exclusion
principle and Coulomb repulsion, the kinetic effects, the electron
density, and an internal magnetic field component.  The `sources' of
these fields are quantum-mechanical in that they are expectation
values of Hermitian operators taken with respect to the system wave
function $\Psi$.  Hence, the perspective hews to the probabilistic
interpretation of quantum mechanics.  The fields, as obtained from
their respective quantal sources, obey equations of classical
physics.  Therefore, as in classical physics, these fields pervade
all space.  The description of the quantum mechanical system in
terms of these 'classical' fields then makes it tangible in the
classical sense. In this context, the fields are determinate.

One significant feature of the law is that in addition to the
expected external electrostatic and Lorentz fields, each electron
also experiences an internal field.  And that these fields are
representative of the intrinsic properties of the system.  Whilst
one might expect an internal field representative of Coulombic and
Pauli principle electron-electron repulsion, one learns that there
exist other components of the internal field. Hence, there exists a
field representative of kinetic effects, and one representative of
the electron density. And, that there also exists an internal
magnetic field component.

The magnetic field contributions to the `Quantal Newtonian' first
law are the external Lorentz and internal magnetic fields.  Provided
the sum of these fields is conservative, i.e. curl-free, then it is
possible to define a scalar (path-independent) magnetic potential
$v_{m} ({\bf{r}})$ in a manner similar to the external scalar
electrostatic potential $v ({\bf{r}})$.  There are other facets of
the quantum system that emerge as a consequence of the `Quantal
Newtonian' first law, and these will be discussed in the text.

The `Quantal Newtonian' first law is a sum rule.  As such it can be
employed as a constraint applied to approximate wave functions or as
a test of the accuracy of such wave functions.

The non-relativistic Schr{\"o}dinger-Pauli Hamiltonian $\hat{H}$ for
spin  $ \frac{1}{2}$ particles is the sum of the Feynman \cite{30}
kinetic $\hat{T}_{F}$, electron-interaction potential $\hat{W}$, and
external electrostatic potential $\hat{V}$ operators.  In atomic
units (charge of electron $-e, e = \hbar = m = 1$) the Hamiltonian
is
\begin{equation}
\hat{H} = \hat{T}_{F} + \hat{W} + \hat{V},
\end{equation}
where
\begin{equation}
\hat{T}_{F} = \frac{1}{2} \sum_{k} ({\boldsymbol{\sigma}}_{k} \cdot
\hat{\bf{p}}_{k, phys}) ({\boldsymbol{\sigma}}_{k} \cdot
\hat{\bf{p}}_{k, phys}),
\end{equation}
\begin{equation}
\hat{W} = \frac{1}{2} \sideset{}{'}\sum_{k,\ell}
 \frac{1}{|{\bf{r}}_{k} - {\bf{r}}_{\ell}|},
\end{equation}
\begin{equation}
\hat{V} = \sum_{k} v ({\bf{r}}_{k}).
\end{equation}
Here the physical momentum operator $\hat{\bf{p}}_{phys} =
(\hat{\bf{p}} + \frac{1}{c} {\bf{A}} ({\bf{r}}))$, with
$\hat{\bf{p}} = - i {\boldsymbol{\nabla}}$ the canonical momentum
operator.  The ${\boldsymbol{\sigma}}$ is the Pauli spin matrix:
${\bf{s}} = \frac{1}{2} {\boldsymbol{\sigma}}$, with ${\bf{s}}$ the
electron spin angular momentum vector operator.  On substituting for
$\hat{\bf{p}}_{phys}$ and ${\boldsymbol{\sigma}}$ in the kinetic
energy operator equation, the Hamiltonian $\hat{H}$ may be written
as
\begin{equation}
\hat{H}  = \frac{1} {2} \sum_{k} \big( \hat{\bf{p}}_{k} +
\frac{1}{c} {\bf{A}} ({\bf{r}}_{k}) \big)^{2} + \frac{1}{c} \sum_{k}
{\boldsymbol{\cal{B}}} ({\bf{r}}_{k}) \cdot {\bf{s}}_{k} + \hat{W} +
\hat{V},
\end{equation}
which then indicates the interaction of the magnetic field with both
the orbital and spin moment of the electrons.  (The former
interaction becomes evident for the case of a uniform magnetic
field.  In the symmetric gauge ${\bf{A}} ({\bf{r}}_{k}) =
\frac{1}{2} {\bf{B}} \times {\bf{r}}_{k}$, the ${\bf{A}}
({\bf{r}}_{k}) \cdot \hat{\bf{p}}_{k}$ term of the Hamiltonian may
be written as $\frac{1}{2} {\bf{B}} \cdot {\bf{L}}_{k}$, with
${\bf{L}}_{k} = {\bf{r}}_{k} \times \hat{\bf{p}}_{k}$ the orbital
angular momentum operator.  In the symmetric gauge,
${\boldsymbol{\nabla}} \cdot {\bf{A}} ({\bf{r}}_{k}) = 0$.  Hence,
in the Hamiltonian, the $\hat{\bf{p}}_{k} \cdot {\bf{A}}
({\bf{r}}_{k})$ term vanishes.  The interaction of the magnetic
field with the spin moment was originally added \emph{ad hoc} to the
Schr{\"o}dinger equation by Pauli.)  It is interesting to note that
both interactions arise via the Feynman kinetic energy operator
$T_{F}$. The spin magnetic moment obtained this way has the correct
gyromagnetic ratio $g=2$. This then is the non-relativistic
derivation of the Schr{\"o}dinger-Pauli Hamiltonian.  The
Hamiltonian $\hat{H}$ of Eq. (5) may also be derived \cite{30} as
the non-relativistic limit of Dirac theory. The corresponding
Schr{\"o}dinger-Pauli equation is then
\begin{equation}
\hat{H}  \Psi ({\bf{X}}) = E \Psi ({\bf{X}}),
\end{equation}
with $\{  \Psi ({\bf{X}}), E \}$ the eigenfunctions and eigenvalues;
${\bf{X}} = {\bf{x}}, \ldots, {\bf{x}}_{N}$; ${\bf{x}} = {\bf{r}}
\sigma$; and ${\bf{r}} \sigma$ the spatial and spin coordinates.

There are three components to the paper:

\setlength{\parindent}{10ex}

1.  The first is comprised of the description of the
quantum-mechanical system as defined by the Hamiltonian $\hat{H}$ of
Eq. (5) in terms of the `classical' fields that satisfy the
corresponding `Quantal Newtonian' first law.  This description is
valid for \emph{arbitrary} state whether ground, excited or
degenerate. (The derivation of the law is given in the Appendix.)
The description leads to physical insights and understandings, not
previously known of the quantum system, and these are then
explained.

2.  The second is a generalization of the Schr{\"o}dinger-Pauli
equation which shows the Hamiltonian $\hat{H}$ to be a functional of
the wave function $\Psi$, i.e. $\hat{H} = \hat{H} [\Psi]$.  Hence,
the Schr{\"o}dinger-Pauli equation can be written in a more general
form as
\begin{equation}
\hat{H} [\Psi] \Psi ({\bf{X}}) = E [\Psi] \Psi ({\bf{X}}).
\end{equation}
In Eq. (7) the fact that the eigenvalues $E$ too are functionals of
the wave function $\Psi$ is also explicitly indicated.  The
generalization of the Schr{\"o}dinger-Pauli equation is a
consequence of the `Quantal Newtonian' first law. It is therefore
valid for arbitrary state.  As the first law is in terms of fields
whose sources are quantum-mechanical expectations of Hermitian
operators taken with respect to the wave function $\Psi$, the
functional $\hat{H} [\Psi]$ is \emph{exactly known}.  The functional
$\hat{H} [\Psi]$ is also \emph{universal} in that it is valid for
any electronic system.  It is evident from the generalized form of
the Schr{\"o}dinger-Pauli equation that it is intrinsically a
self-consistent eigenvalue equation.  In the self-consistent
procedure, the eigenvalue $E$ at each iteration depends upon the
solution of the equation for that iteration.  It is for this reason
that the eigenvalue $E$ is written as the functional $E[\Psi]$. (The
generalized Schr{\"o}dinger-Pauli equation is akin to the
Hartree-Fock theory \cite{31,32} equations in which the
corresponding Hamiltonian $\hat{H}^\mathrm{HF}$ is a functional of
the single-particle spin-orbitals $\phi_{i} ({\bf{x}})$, i.e.
$\hat{H}^\mathrm{HF} = \hat{H}^\mathrm{HF} [\phi_{i} ({\bf{x}})]$.
The Hartree-Fock theory equations are then $\hat{H}^\mathrm{HF}
[\phi_{i}] \phi_{i} ({\bf{x}}) = \epsilon_{i} \phi_{i} ({\bf{x}})$,
where the $\epsilon_{i}$ are the eigenvalues. These equations are
then solved self-consistently \cite{33}. Such self-consistent
equations also arise within all local effective potential theories
such as the Optimized Potential Method \cite{34,35,36}, Kohn-Sham
\cite{37}, and quantal \cite{2,3,4} density functional theories.)

3.  The third component of the paper constitutes the mapping of the
interacting system of electrons as defined by the
Schr{\"o}dinger-Pauli equation of Eq. (6) to one of noninteracting
fermions possessing the same \emph{basic variables} of the density
$\rho ({\bf{r}})$ and physical current density ${\bf{j}}
({\bf{r}})$. The further constraints of the mapping are that of
fixed electron number $N$, and total orbital ${\bf{L}}$ and spin
${\bf{S}}$ angular momentum.  (Basic variables in quantum mechanics
are gauge invariant properties, knowledge of which uniquely
determine the external scalar and vector potentials to within a
constant and gradient of a scalar function, respectively.) The
mapping is accomplished via quantal density functional theory
(QDFT).  The mapping is valid for arbitrary state of the interacting
system.  The state of the model system is also arbitrary provided
the constraints are satisfied. The reasons for this mapping are
twofold:

{\leftskip=2cm\relax
 \rightskip=2cm\relax
\noindent (a) The mapping to the model system allows for the
determination of properties of the quantum system not obtainable
solely via the solution of the Schr{\"o}dinger-Pauli equation.  Such
a property is the contribution of electron correlations due to the
Pauli exclusion principle and Coulomb repulsion to the kinetic
energy -- the correlation-kinetic energy. Further, as a consequence
of the mapping, it is also possible to separate the contributions to
the total energy of the correlations due to the Pauli principle and
Coulomb repulsion. (The solution $\Psi ({\bf{X}})$ of the
Schr{\"o}dinger-Pauli Eq. (6,7) accounts for both types of electron
correlations, but they are not separable. In quantum chemistry, the
separation is accomplished in an approximate manner by performing a
Hartree-Fock theory calculation which then leads to the exchange
energy -- the contribution due to the correlations arising from the
Pauli principle.  But this latter model differs from the original
fully-interacting system as its density $\rho ({\bf{r}})$ and
physical current density ${\bf{j}} ({\bf{r}})$ are different.  In
the QDFT mapping, the model system density and physical current
density are the same as that of the interacting electrons.)  It is
also possible to determine the ionization potential via the mapping
to the model system of noninteracting fermions.  The highest
occupied eigenvalue of the corresponding differential equation  is
the negative of the ionization potential.  (It requires two separate
energy calculations to determine the ionization potential within
Schr{\"o}dinger-Pauli theory: one for the charge-neutral and the
other for the ionized system.)\\
 \par}

{\leftskip=2cm\relax
 \rightskip=2cm\relax
\noindent (b) The equations governing the model system of
noninteracting fermions is easier to solve.  The corresponding `wave
function' is a Slater determinant of the model fermion
spin-orbitals.  The QDFT mapping provides the precise
\emph{physical} definition of the local effective potential in which
all the many-body effects are incorporated.  It is the work done by
the model fermion in a conservative effective field. This potential
then generates the interacting system density $\rho ({\bf{r}})$ and
physical current density ${\bf{j}} ({\bf{r}})$. \\
 \par}

{\leftskip=2cm\relax
 \rightskip=2cm\relax
\noindent The rationale for the choice of the densities $\{ \rho
({\bf{r}}), {\bf{j}} ({\bf{r}}) \}$ as the basic variables in the
mapping stems from the ground state theorem of Hohenberg-Kohn (HK)
\cite{38} and of its extension by Pan-Sahni (PS) \cite{39} to the
presence of a uniform magnetic field. For a system of $N$ electrons
in an external electrostatic field ${\boldsymbol{\cal{E}}}
({\bf{r}}) = - {\boldsymbol{\nabla}} v({\bf{r}})/e$, and in a
nondegenerate ground state, HK proved that knowledge of the ground
state density $\rho ({\bf{r}})$ uniquely determined the external
scalar potential $v({\bf{r}})$ to within a constant. The constraint
in the proof is that of fixed electron number $N$.  As the kinetic
$\hat{T}$ and electron-interaction $\hat{W}$ potential operators are
assumed known, so thus is the Hamiltonian. Solution of the
Schr{\"o}dinger equation then leads to the eigenfunctions and
eigenvalues of the system. Hence, the nondegenerate ground state
density $\rho ({\bf{r}})$ constitutes a basic variable.  What PS
proved was that in the added presence of a \emph{uniform}
magnetostatic field ${\boldsymbol{\cal{B}}} ({\bf{r}}) =
{\boldsymbol{\nabla}} \times {\bf{A}} ({\bf{r}})$, knowledge of the
nondegenerate ground state $\{ \rho ({\bf{r}}), {\bf{j}} ({\bf{r}})
\}$ uniquely determines the potentials $\{ v({\bf{r}}), {\bf{A}}
({\bf{r}}) \}$ to within a constant and gradient of a scalar
function, respectively. The constraints in the proof are that of
fixed electron number $N$, orbital ${\bf{L}}$, and spin ${\bf{S}}$
angular momentum.  The PS proof was for both spinless electrons and
electrons with spin. Again, with the Hamiltonian now known, the
solution of the corresponding Schr{\"o}dinger and
Schr{\"o}dinger-Pauli equations then leads to the system
eigenfunctions and eigenvalues. Hence, in the presence of a
magnetostatic field, the nondegenerate ground state $\{ \rho
({\bf{r}}), {\bf{j}} ({\bf{r}}) \}$ constitute basic variables. (The
HK and PS proofs differ. There is a fundamental reason for this.  In
HK the relationship between $v ({\bf{r}})$ and the nondegenerate
ground state $\Psi$ is proved to be bijective or one-to-one. In the
presence of a magnetic field, however, the relationship between $\{
v({\bf{r}}), {\bf{A}} ({\bf{r}}) \}$ and the nondegenerate ground
state $\Psi$ is many-to-one and can be infinite-to-one.  PS
explicitly account for this many-to-one relationship, and in doing
so, the proof follows a different path.) The theorems of HK and PS
are ground state theorems.  Thus, within HK, the mapping is from an
interacting system in a ground state to one of noninteracting
fermions also in a ground state possessing the same density $\rho
({\bf{r}})$. This is the mapping performed, for example, in
Kohn-Sham density functional theory.  However, within QDFT, the
mapping to the model system with the same $\rho ({\bf{r}})$ or $\{
\rho ({\bf{r}}), {\bf{j}} ({\bf{r}}) \}$ is possible for ground,
excited, and degenerate states of the interacting system
\cite{18,19,40,41,42,43}. \\
 \par}

To elucidate the ideas underlying the quantal-source--field
perspective, the satisfaction of the `Quantal Newtonian' first law,
and the intrinsic self-consistent nature of the
Schr{\"o}dinger-Pauli equation, we apply them to the first excited
triplet $2^{3} S$ state of a quantum dot in a magnetic field in the
following paper \cite{44}.  The present paper on
Schr{\"o}dinger-Pauli theory is a generalization of work on the
Schr{\"o}dinger theory of electrons \cite{45,46,47}.  As such the
description of Schr{\"o}dinger theory within this new perspective
constitutes a special case.

In Sect. II we present the quantal source-field perspective of the
Schr{\"o}dinger-Pauli theory, and describe the new physical insights
as obtained from the `Quantal Newtonian' first law.  The
generalization of the Schr{\"o}dinger-Pauli equation to exhibit its
self-consistent nature is discussed in Sect. III.  In Sect. IV the
local effective potential quantal density functional theory
description of Schr{\"o}dinger-Pauli theory is developed.  Finally,
in Sect. V, we summarize the conclusions of the work and propose
further generalizations to the time-dependent Schr{\"o}dinger-Pauli
theory, and to relativistic quantum mechanics via the Dirac
equation.

\section{Description in Terms of Quantal Sources and Fields:
The `Quantal Newtonian' First Law}

In this section the quantum-mechanical system defined by the
Schr{\"o}dinger-Pauli Hamiltonian is described in terms of
'classical' fields as experienced by each electron.  These fields
arise from quantal sources that are expectation values of Hermitian
operators, or of complex operators whose real and imaginary parts
are Hermitian, taken with respect to the system wave function $\Psi
({\bf{X}})$. Knowledge of the structure of the quantal sources is
then predictive of the structure of the corresponding fields. The
fields satisfy the `Quantal Newtonian' first law -- the equation of
motion of the individual electron.  The description is valid for
arbitrary state.  Further, the total energy $E$ of the system, and
its components can also be expressed in integral virial form in
terms of these fields. The fields can be separated into two
categories: an external ${\boldsymbol{\cal{F}}}^\mathrm{ext}
({\bf{r}})$ and an internal ${\boldsymbol{\cal{F}}}^\mathrm{int}
({\bf{r}})$ field.  To define these fields, the Hamiltonian
$\hat{H}$ of Eq. (5) is rewritten as
\begin{eqnarray}
\hat{H} = \hat{T} + \frac{1}{c} \int \hat{\bf{j}}_{p} ({\bf{r}})
\cdot {\bf{A}} ({\bf{r}}) d {\bf{r}} &+& \frac{1}{2c} \int
\hat{\bf{j}}_{d} ({\bf{r}}) \cdot {\bf{A}} ({\bf{r}}) d {\bf{r}}
\nonumber\\
&+& \frac{1}{c} \int \hat{\bf{j}}_{m} ({\bf{r}}) \cdot {\bf{A}}
({\bf{r}}) d {\bf{r}} + \hat{W} + \hat{V}.
\end{eqnarray}
In the above equation, $\hat{T}$ is the canonical kinetic energy
operator:
\begin{equation}
\hat{T} = \frac{1}{2} \sum_{k} \hat{p}_{k}^{2},
\end{equation}
and where the paramagnetic $\hat{\bf{j}}_{p} ({\bf{r}})$,
diamagnetic $\hat{\bf{j}}_{d} ({\bf{r}})$, and magnetization
$\hat{\bf{j}}_{m} ({\bf{r}})$ current density operators are defined
as
\begin{equation}
\hat{\bf{j}}_{p} ({\bf{r}}) = \frac{1}{2} \sum_{k} \big[
\hat{\bf{p}}_{k} \delta({\bf{r}}_{k} - {\bf{r}}) +
\delta({\bf{r}}_{k} - {\bf{r}}) \hat{\bf{p}}_{k} \big],
\end{equation}
\begin{equation}
\hat{\bf{j}}_{d} ({\bf{r}}) = \frac{1}{c} \hat{\rho} ({\bf{r}})
{\bf{A}} ({\bf{r}}),
\end{equation}
and
\begin{equation}
\hat{\bf{j}}_{m} ({\bf{r}}) = - c {\boldsymbol{\nabla}} \times
\hat{\bf{m}} ({\bf{r}}).
\end{equation}
In turn the electronic density $\hat{\rho} ({\bf{r}})$ and
magnetization density $\hat{\bf{m}} ({\bf{r}})$ operators of these
equations are defined as
\begin{equation}
\hat{\rho} ({\bf{r}}) = \sum_{k} \delta ({\bf{r}}_{k} - {\bf{r}}),
\end{equation}\
and
\begin{equation}
\hat{\bf{m}} ({\bf{r}}) = - \frac{1}{c} \sum_{k} {\bf{s}}_{k} \delta
({\bf{r}}_{k} - {\bf{r}}).
\end{equation}
The physical current density operator $\hat{\bf{j}} ({\bf{r}})$ is
then obtained via its definition \cite{48} as
\begin{equation}
\hat{\bf{j}} ({\bf{r}}) = c \frac{\partial \hat{H}} {\partial
{\bf{A}} ({\bf{r}})} = \hat{\bf{j}}_{p} ({\bf{r}}) +
\hat{\bf{j}}_{d} ({\bf{r}}) + \hat{\bf{j}}_{m} ({\bf{r}}).
\end{equation}
In terms of the current density $\hat{\bf{j}} ({\bf{r}})$, the
Hamiltonian $\hat{H}$ of Eq. (8) may be written as
\begin{equation}
\hat{H} = \hat{T} + \frac{1}{c} \int \hat{\bf{j}} ({\bf{r}}) \cdot
{\bf{A}} ({\bf{r}}) d {\bf{r}} - \frac{1} {2c^{2}} \int \hat{\rho}
({\bf{r}}) A^{2} ({\bf{r}}) d {\bf{r}}+ \hat{W} + \hat{V},
\end{equation}
which then emphasizes the significance of both the electronic and
physical current densities to the quantum system.

\subsection{\emph{External Field ${\boldsymbol{\cal{F}}}^\mathrm{ext}
({\bf{r}})$}}

The external field ${\boldsymbol{\cal{F}}}^\mathrm{ext} ({\bf{r}})$
experienced by each electron is the sum of the binding electrostatic
${\boldsymbol{\cal{E}}} ({\bf{r}})$ and Lorentz
${\boldsymbol{\cal{L}}} ({\bf{r}})$ fields:
\begin{equation}
{\boldsymbol{\cal{F}}}^\mathrm{ext} ({\bf{r}}) =
{\boldsymbol{\cal{E}}} ({\bf{r}}) - {\boldsymbol{\cal{L}}}
({\bf{r}}) = - {\boldsymbol{\nabla}} v ({\bf{r}}) -
{\boldsymbol{\cal{L}}} ({\bf{r}}),
\end{equation}
where the Lorentz field ${\boldsymbol{\cal{L}}} ({\bf{r}})$ is
defined in terms of the Lorentz `force' ${\boldsymbol{\ell}}
({\bf{r}})$ and electronic density $\rho ({\bf{r}})$ (charge) as
\begin{equation}
{\boldsymbol{\cal{L}}} ({\bf{r}}) = \frac{{\boldsymbol{\ell}}
({\bf{r}})} {\rho ({\bf{r}})},
\end{equation}
with
\begin{equation}
{\boldsymbol{\ell}} ({\bf{r}}) = {\bf{j}} ({\bf{r}}) \times
{\boldsymbol{\cal{B}}} ({\bf{r}}).
\end{equation}
The electronic $\rho ({\bf{r}})$ and physical current ${\bf{j}}
({\bf{r}})$ densities are, respectively, the expectation values of
the operators $\hat{\rho} ({\bf{r}})$ and $\hat{\bf{j}} ({\bf{r}})$:
\begin{equation}
\rho ({\bf{r}}) = \langle \Psi ({\bf{X}}) | \hat{\rho} ({\bf{r}}) |
\Psi ({\bf{X}}) \rangle,
\end{equation}
and
\begin{equation}
{\bf{j}} ({\bf{r}}) = \langle \Psi ({\bf{X}}) | \hat{\bf{j}}
({\bf{r}}) | \Psi ({\bf{X}}) \rangle.
\end{equation}

\subsection{\emph{Internal Field ${\boldsymbol{\cal{F}}}^\mathrm{int}
({\bf{r}})$}}

The internal field ${\boldsymbol{\cal{F}}}^\mathrm{int} ({\bf{r}})$
is a sum of components each descriptive of a property of the system:
an electron-interaction field ${\boldsymbol{\cal{E}}}_\mathrm{ee}
({\bf{r}})$ representative of electron correlations  due to the
Pauli exclusion principle and Coulomb repulsion; a kinetic field
${\boldsymbol{\cal{Z}}} ({\bf{r}})$ from which the kinetic energy
density and kinetic energy can be obtained; the differential density
field ${\boldsymbol{\cal{D}}} ({\bf{r}})$ representative of the
electron density; and finally an internal magnetic field component
${\boldsymbol{\cal{I}}}_{m} ({\bf{r}})$.  Thus,
\begin{equation}
{\boldsymbol{\cal{F}}}^\mathrm{int} ({\bf{r}}) =
{\boldsymbol{\cal{E}}}_\mathrm{ee} ({\bf{r}}) -
{\boldsymbol{\cal{Z}}} ({\bf{r}}) - {\boldsymbol{\cal{D}}}
({\bf{r}}) - {\boldsymbol{\cal{I}}}_{m} ({\bf{r}}).
\end{equation}
The component fields and their respective quantal sources are
defined next.

The electron-interaction field ${\boldsymbol{\cal{E}}}_\mathrm{ee}
({\bf{r}})$ in terms of the electron-interaction `force'
${\bf{e}}_\mathrm{ee} ({\bf{r}})$ and density $\rho ({\bf{r}})$
(charge) is
\begin{equation}
{\boldsymbol{\cal{E}}}_\mathrm{ee} ({\bf{r}}) =
\frac{{\bf{e}}_\mathrm{ee} ({\bf{r}})} {\rho ({\bf{r}})},
\end{equation}
where ${\bf{e}}_\mathrm{ee} ({\bf{r}})$ is obtained via Coulomb's
law from its \emph{nonlocal} (dynamic) quantal source, the
pair-correlation function $P ({\bf{ r r}}')$:
\begin{equation}
{\bf{e}}_\mathrm{ee} ({\bf{r}}) = \int \frac{P ({\bf{ r r}}')
({\bf{r - r}}')} {| {\bf{r - r}}'|^{3}} d {\bf{r}}'.
\end{equation}
with $P ({\bf{ r r}}')$ the expectation value
\begin{equation}
P ({\bf{ r r}}') = \langle \Psi ({\bf{X}}) | \hat{P} ({\bf{r r}}') |
\Psi ({\bf{X}}) \rangle,
\end{equation}
of the pair-correlation operator
\begin{equation}
\hat{P} ({\bf{ r r}}') = \sideset{}{'}\sum_{k,\ell} \delta
({\bf{r}}_{k} - {\bf{r}}) \delta ({\bf{r}}_{\ell} - {\bf{r}}').
\end{equation}
The electron-interaction field ${\boldsymbol{\cal{E}}}_\mathrm{ee}
({\bf{r}})$ may equivalently be thought of as arising via Coulomb's
law from the quantal source of the pair-correlation density $g ({\bf
{rr}}') = P ({\bf {rr}}')/\rho ({\bf{r}})$.  The pair-correlation
density can be separated into its \emph{local}  $\rho ({\bf{r}}')$
and \emph{nonlocal} $\rho_{xc} ({\bf{r r}}')$ components: $g ({\bf
{rr}}') = \rho ({\bf{r}}') + \rho_{xc} ({\bf{r r}}')$, where
$\rho_{xc} ({\bf{r r}}')$ is the quantum-mechanical Fermi-Coulomb
hole charge distribution.  Thus, the field
${\boldsymbol{\cal{E}}}_\mathrm{ee} ({\bf{r}})$ may be written as a
sum of its Hartree ${\boldsymbol{\cal{E}}}_{H} ({\bf{r}})$ and
Pauli-Coulomb ${\boldsymbol{\cal{E}}}_\mathrm{xc} ({\bf{r}})$
components:
\begin{equation}
{\boldsymbol{\cal{E}}}_\mathrm{ee} ({\bf{r}}) =
{\boldsymbol{\cal{E}}}_{H} ({\bf{r}}) +
{\boldsymbol{\cal{E}}}_\mathrm{xc} ({\bf{r}}),
\end{equation}
where
\begin{equation}
{\boldsymbol{\cal{E}}}_{H} ({\bf{r}}) = \int \frac{\rho ({\bf{r}}')
({\bf{ r-r}}')} {|{\bf{r-r}}'|^{3}} d {\bf{r}}',
\end{equation}
and
\begin{equation}
{\boldsymbol{\cal{E}}}_\mathrm{xc} ({\bf{r}}) = \int
\frac{\rho_\mathrm{xc}  ({\bf{r r}}') ({\bf{ r-r}}')}
{|{\bf{r-r}}'|^{3}} d {\bf{r}}'.
\end{equation}
Note that in traditional quantum mechanics, it is not possible to
further split the Fermi-Coulomb hole into its Fermi $\rho_{x}
({\bf{r r}}')$ and Coulomb $\rho_{c} ({\bf{r r}}')$ components. In
other words, it is not possible to separate the correlations due to
the Pauli exclusion principle and Coulomb repulsion.  This
separation will be accomplished in Sect. IV via quantal density
functional theory.

The kinetic field ${\boldsymbol{\cal{Z}}} ({\bf{r}})$ is defined in
terms of the kinetic `force' ${\bf{z}} ({\bf{r}})$ and the density
$\rho ({\bf{r}})$ as
\begin{equation}
{\boldsymbol{\cal{Z}}} ({\bf{r}}) = \frac{{\bf{z}} ({\bf{r}})} {\rho
({\bf{r}})}.
\end{equation}
The kinetic `force' is obtained from its \emph{nonlocal} (dynamic)
quantal source, the single-particle density matrix $\gamma ({\bf{r
r}}')$ as follows:
\begin{equation}
z_{\alpha} ({\bf{r}}) = 2 \sum_{\beta} \nabla_{\beta} t_{\alpha
\beta} ({\bf{r}} ; \gamma),
\end{equation}
where the second-rank kinetic energy tensor $t_{\alpha \beta}
({\bf{r}} ; \gamma)$ is
\begin{equation}
t_{\alpha \beta} ({\bf{r}} ; \gamma) = \frac{1}{4} \bigg[
\frac{\partial^{2}} {\partial r'_{\alpha} \partial r''_{\beta}} +
\frac{\partial^{2}} {\partial r'_{\beta} \partial r''_{\alpha}}
\bigg] \gamma ({\bf{r}}' {\bf{r}}'') \bigg|_{{\bf{r}}' = {\bf{r}}''
= r} .
\end{equation}
The quantal source $\gamma ({\bf{r r}}')$is the expectation value
\begin{equation}
\gamma ({\bf{r r}}') = \langle \Psi ({\bf{X}}) | \hat{\gamma}
({\bf{r r}}') | \Psi ({\bf{X}}) \rangle,
\end{equation}
with the complex density matrix operator $\hat{\gamma} ({\bf{r
r}}')$ being \cite{49,50}
\begin{equation}
\hat{\gamma} ({\bf{r r}}')  = \hat{A} + i \hat{B},
\end{equation}
\begin{eqnarray}
\hat{A} = \frac{1}{2} \sum_{k} \big[\delta ({\bf{r}}_{k} - {\bf{r}})
T_{k} ({\bf{a}}) + \delta ({\bf{r}}_{k} - {\bf{r}}') T_{k}
(-{\bf{a}}) \big], \\
\hat{B} = - \frac{i}{2} \sum_{k} \big[\delta ({\bf{r}}_{k} -
{\bf{r}}) T_{k} ({\bf{a}}) - \delta ({\bf{r}}_{k} - {\bf{r}}') T_{k}
(-{\bf{a}}) \big],
\end{eqnarray}
with $T_{k} ({\bf{a}})$ a translation operator such that $T_{k}
({\bf{a}}) \psi ( \ldots {\bf{r}}_{k}, \ldots) = \psi ( \ldots
{\bf{r}}_{k} + {\bf{a}}, \ldots)$ and ${\bf{a}} = {\bf{r}}' -
{\bf{r}}$.  The operators $\hat{A}$ and $\hat{B}$ are each
Hermitian.

The differential density field ${\boldsymbol{\cal{D}}} ({\bf{r}})$
whose quantal source is the \emph{local} electron density $\rho
({\bf{r}})$, is defined in terms of the corresponding `force' $d
({\bf{r}})$ and density $\rho ({\bf{r}})$ as
\begin{equation}
{\boldsymbol{\cal{D}}} ({\bf{r}}) = \frac{d ({\bf{r}})} {\rho
({\bf{r}})},
\end{equation}
where
\begin{equation}
d ({\bf{r}}) = - \frac{1}{4} {\boldsymbol{\nabla}} \nabla^{2} \rho
({\bf{r}}).
\end{equation}

The magnetic field contribution ${\boldsymbol{\cal{I}}}_{m}
({\bf{r}})$ to the internal field in terms of the `force'
${\bf{i}}_{m} ({\bf{r}})$ and the density $\rho ({\bf{r}})$ is
\begin{equation}
{\boldsymbol{\cal{I}}}_{m} ({\bf{r}}) = \frac{{\bf{i}}_{m}
({\bf{r}})} {\rho ({\bf{r}})},
\end{equation}
where
\begin{equation}
i_{m, \alpha} ({\bf{r}}) = \sum_{\beta} \nabla_{\beta} I_{\alpha
\beta} ({\bf{r}}),
\end{equation}
and the second-rank tensor $I_{\alpha \beta} ({\bf{r}})$ is
\begin{equation}
I_{\alpha \beta} ({\bf{r}}) = \big[ j_{\alpha} ({\bf{r}}) A_{\beta}
({\bf{r}}) + j_{\beta} ({\bf{r}}) A_{\alpha} ({\bf{r}}) \big] - \rho
({\bf{r}}) A_{\alpha} ({\bf{r}}) A_{\beta} ({\bf{r}}),
\end{equation}
with ${\bf{j}} ({\bf{r}})$ the quantal source of the field.

The individual components of the internal field
${\boldsymbol{\cal{F}}}^\mathrm{int} ({\bf{r}})$ are in general not
conservative.  However, as shown below, their sum taken together
with the Lorentz field is conservative.  Under conditions of certain
symmetry, the individual components can each be separately
conservative.

\subsection{\emph{`Quantal Newtonian' First Law}}

The equation of motion or `Quantal Newtonian' first law is satisfied
by \emph{each electron} of the physical system defined by the
Schr{\"o}dinger-Pauli equation of Eq. (6).  The law states that the
sum of the external ${\boldsymbol{\cal{F}}}^\mathrm{ext} ({\bf{r}})$
and internal ${\boldsymbol{\cal{F}}}^\mathrm{int} ({\bf{r}})$ fields
experienced by each electron vanishes:
\begin{equation}
{\boldsymbol{\cal{F}}}^\mathrm{ext} ({\bf{r}}) +
{\boldsymbol{\cal{F}}}^\mathrm{int} ({\bf{r}}) = 0.
\end{equation}
The law is derived employing the continuity condition
\begin{equation}
{\boldsymbol{\nabla}} \cdot {\bf{j}} ({\bf{r}}) = 0.
\end{equation}
Thus, the quantal source-field perspective of the
Schr{\"o}dinger-Pauli theory is consistent with Schr{\"o}dinger's
\cite{51} insight that satisfaction of this condition is the
explanation of the lack of radiation in a stationary state.  The
`Quantal Newtonian' first law is valid for arbitrary state.  It is
also gauge invariant.

\subsection{\emph{Total Energy and Components}}

The terms of the total energy $E$ -- the canonical kinetic $T$, the
electron-interaction $E_\mathrm{ee}$, and its Hartree $E_{H}$ and
Pauli-Coulomb $E_{xc}$ components, -- can each be expressed in
integral virial form in terms of the corresponding fields
${\boldsymbol{\cal{Z}}} ({\bf{r}})$,
${\boldsymbol{\cal{E}}}_\mathrm{ee} ({\bf{r}})$,
${\boldsymbol{\cal{E}}}_{H} ({\bf{r}})$,
${\boldsymbol{\cal{E}}}_{xc} ({\bf{r}})$. With the exception of
${\boldsymbol{\cal{E}}}_{H} ({\bf{r}})$ which is conservative, these
expressions are valid irrespective of whether the fields are
conservative.  Thus,
\begin{eqnarray}
T &=& - \frac{1}{2} \int \rho ({\bf{r}}) {\bf{r}} \cdot
{\boldsymbol{\cal{Z}}} ({\bf{r}}) d {\bf{r}}, \\
E_\mathrm{ee} &=& \int \rho ({\bf{r}}) {\bf{r}} \cdot
{\boldsymbol{\cal{E}}}_\mathrm{ee} ({\bf{r}}) d {\bf{r}}, \\
E_{H} &=& \int \rho ({\bf{r}}) {\bf{r}} \cdot
{\boldsymbol{\cal{E}}}_{H} ({\bf{r}}) d {\bf{r}}, \\
E_{xc} &=& \int \rho ({\bf{r}}) {\bf{r}} \cdot
{\boldsymbol{\cal{E}}}_{xc} ({\bf{r}}) d {\bf{r}}.
\end{eqnarray}
The contribution of the conservative external electrostatic field
${\boldsymbol{\cal{E}}} ({\bf{r}}) = -{\boldsymbol{\nabla}} v
({\bf{r}})$ to the energy $E_{el}$ can be written directly in terms
of the potential $v ({\bf{r}})$ as
\begin{equation}
E_{es} = \int \rho ({\bf{r}}) v ({\bf{r}}) d {\bf{r}}.
\end{equation}
Note that $v ({\bf{r}})$ is path-independent.  The energy can also
be written in integral virial form, but the coefficient of the
expression depends upon the degree of the homogeneous function $v
({\bf{r}})$. Hence, for the Coulombic potential for which the degree
is $-1$, the expression is
\begin{equation}
E_{es} = \int \rho ({\bf{r}}) {\bf{r}} \cdot {\boldsymbol{\cal{E}}}
({\bf{r}}) d {\bf{r}}.
\end{equation}
For the magnetic field contribution to the energy, \emph{i.e}. the
contribution of the Lorentz ${\boldsymbol{\cal{L}}} ({\bf{r}})$ and
internal magnetic ${\boldsymbol{\cal{I}}}_{m} ({\bf{r}})$ field
components, define the field
\begin{equation}
{\boldsymbol{\cal{M}}} ({\bf{r}}) = - [{\boldsymbol{\cal{L}}}
({\bf{r}}) + {\boldsymbol{\cal{I}}}_{m} ({\bf{r}})] .
\end{equation}
If the field ${\boldsymbol{\cal{M}}} ({\bf{r}})$ is conservative,
\emph{i.e}. ${\boldsymbol{\nabla}} \times {\boldsymbol{\cal{M}}}
({\bf{r}}) = 0$, then one can define a magnetic scalar potential
$v_{m} ({\bf{r}})$ as
\begin{equation}
{\boldsymbol{\cal{M}}} ({\bf{r}}) = - {\boldsymbol{\nabla}} v_{m}
({\bf{r}}).
\end{equation}
This implies that $v_{m} ({\bf{r}})$ is path-independent.  The
magnetic contribution $E_\mathrm{mag}$ to the energy is then
\begin{equation}
E_\mathrm{mag} = \int \rho ({\bf{r}}) v_{m} ({\bf{r}}) d {\bf{r}}.
\end{equation}
The $E_\mathrm{mag}$ can also be written in integral virial form
depending on the degree of the homogeneous function $v_{m}
({\bf{r}})$.  If $v_{m} ({\bf{r}})$ is of degree $2$ as for the
harmonic oscillator, then
\begin{equation}
E_\mathrm{mag} = - \frac{1}{2} \int \rho ({\bf{r}}) {\bf{r}} \cdot
{\boldsymbol{\cal{M}}} ({\bf{r}}) d {\bf{r}}.
\end{equation}
In the general case when ${\boldsymbol{\nabla}} \times
{\boldsymbol{\cal{M}}} ({\bf{r}}) \neq 0$, the expression is
\begin{equation}
E_\mathrm{mag} = \int \rho ({\bf{r}}) {\bf{r}} \cdot
{\boldsymbol{\cal{M}}} ({\bf{r}}) d {\bf{r}}.
\end{equation}
The total energy $E$ may then be expressed as
\begin{eqnarray}
E &=& T + E_\mathrm{ee} + E_{es} + E_\mathrm{mag} \\
&=& T + E_{H} + E_{xc} + E_{es} + E_\mathrm{mag}.
\end{eqnarray}

It is evident from the above that the quantum-mechanical system
defined via the Schr{\"o}dinger-Pauli equation can be alternatively
described from the perspective of the individual electron.  This
description is in terms of `classical' fields experienced by each
electron, with the fields arising from quantal sources.  The fields
satisfy the `Quantal Newtonian' first law or equation of motion for
each electron.  The total energy $E$ and its components can also be
expressed in terms of these fields.

\subsection{\emph{Further Physical and Mathematical Insights}}

In addition to the above new perspective, further understandings of
the Schr{\"o}dinger-Pauli system may be gleaned from the `Quantal
Newtonian' first law.  These are as follows:

\textbf{(\emph{i})}   The Hamiltonian of a system of classical
particles in an electrostatic and magnetostatic field contains both
a scalar and vector potential representative respectively of these
fields. From the correspondence principle, these same potentials
appear in the quantum-mechanical Hamiltonian. Hence, it is
understood that each electron of the quantum system in such fields
experiences a force due to the electrostatic field, and a Lorentz
force due to the magnetic field.   Whilst the electrostatic force is
explicit via the scalar potential, the Lorentz force does not appear
explicitly in the quantum-mechanical Hamiltonian.   The `Quantal
Newtonian' first law now makes the existence of both forces acting
on each electron \emph{explicit} via the external field
${\boldsymbol{\cal{F}}}^\mathrm{ext} ({\bf{r}})$ which is the sum of
the electrostatic ${\boldsymbol{\cal{E}}} ({\bf{r}})$ and Lorentz
${\boldsymbol{\cal{L}}} ({\bf{r}})$ fields, the latter involving the
Lorentz force.

\textbf{(\emph{ii})}  As is the case for classical particles
interacting via Newton's third law forces, and the resulting
Newton's first law for each particle, each electron of the quantum
system is observed via the `Quantal Newtonian' first law to also
experience an internal field ${\boldsymbol{\cal{F}}}^\mathrm{int}
({\bf{r}})$.  The components of this field are representative of
fundamental properties of the quantum system: electron correlations
due to the Pauli exclusion principle and Coulomb repulsion,
${\boldsymbol{\cal{E}}}_\mathrm{ee} ({\bf{r}})$; kinetic effects,
${\boldsymbol{\cal{Z}}} ({\bf{r}})$; electron density,
${\boldsymbol{\cal{D}}} ({\bf{r}})$; and an internal magnetic field
component, ${\boldsymbol{\cal{I}}}_{m} ({\bf{r}})$.  The existence
of the internal field ${\boldsymbol{\cal{F}}}^\mathrm{int}
({\bf{r}})$ and of its property related components would be unknown
but for the `Quantal Newtonian' first law.

\textbf{(\emph{iii})}  In summing the `Quantal Newtonian' first law
over all the electrons, the contribution of the internal field
${\boldsymbol{\cal{F}}}^\mathrm{int} ({\bf{r}})$ vanishes, leading
to Ehrenfest's \cite{52} theorem for a stationary state:  $\int \rho
({\bf{r}}) {\boldsymbol{\cal{F}}}^\mathrm{ext} ({\bf{r}}) d {\bf{r}}
= 0$.

\textbf{(\emph{iv})} The external scalar potential $v ({\bf{r}})$
which appears in the quantum-mechanical Hamiltonian represents the
potential energy of each electron in the presence of the field of
the positively charged nucleus in atoms, molecules, and solids.  It
could represent the potential due to the field of the positive
jellium background model of solids (metals) employed to study the
uniform electron gas or the study of the metal-vacuum interface
\cite{3,53}, or the fractional quantum Hall effect \cite{9,10}. The
potential, furthermore, is path-independent.  The `Quantal
Newtonian' first law, however, provides a deeper physical
understanding of this potential in terms of the properties of the
system. Further, it affords an interpretation of the potential in
the rigorous classical sense. It follows from the `Quantal
Newtonian' first law of Eq. (41) that \emph{the potential $v
({\bf{r}})$ is the work done to move an electron from some reference
point at infinity to its position at ${\bf{r}}$ in the force of a
conservative field ${\boldsymbol{\cal{F}}} ({\bf{r}})$}:
\begin{equation}
v ({\bf{r}}) = \int^{{\bf{r}}}_{\infty} {\boldsymbol{\cal{F}}}
({\bf{r}}') \cdot d {\boldsymbol{\ell}}',
\end{equation}
where ${\boldsymbol{\cal{F}}} ({\bf{r}}) =
{\boldsymbol{\cal{F}}}^\mathrm{int} ({\bf{r}}) -
{\boldsymbol{\cal{L}}} ({\bf{r}}) =
{\boldsymbol{\cal{E}}}_\mathrm{ee} ({\bf{r}}) -
{\boldsymbol{\cal{Z}}} ({\bf{r}}) - {\boldsymbol{\cal{D}}}
({\bf{r}}) - {\boldsymbol{\cal{I}}}_{m} ({\bf{r}}) -
{\boldsymbol{\cal{L}}} ({\bf{r}})$. As the field
${\boldsymbol{\cal{F}}} ({\bf{r}})$ is conservative, the
${\boldsymbol{\nabla}} \times {\boldsymbol{\cal{F}}} ({\bf{r}}) =
0$. Hence, the work done is \emph{path-independent}, and therefore
$v ({\bf{r}})$ constitutes a potential energy. It is reiterated that
the `Quantal Newtonian' first law is valid for arbitrary state.
Hence, the potential function $v ({\bf{r}})$ as expressed in Eq.
(57) remains the same irrespective of the state of the system.

\textbf{(\emph{v})}  In the Hamiltonian of Eq. (6), the potential
energy function $v ({\bf{r}})$ binding the electrons is assumed
analytically known. It could be Coulombic $(-Ze^{2}/r)$, harmonic
$(\frac{1}{2} kr^{2})$, screened-Coulomb Yukawa $(-Ze^{2}
exp(-\lambda r/r)$, etc. The `Quantal Newtonian' first law written
as in Eq. (57) then shows that this analytical function $v
({\bf{r}})$ depends on all the components of the internal field
${\boldsymbol{\cal{F}}}^\mathrm{int} ({\bf{r}})$ of the system and
the Lorentz field ${\boldsymbol{\cal{L}}} ({\bf{r}})$. Thus, the
potential $v ({\bf{r}})$ is inherently related to and constructed
via the properties of the system. Further, if the various internal
fields are separately conservative, then the function $v ({\bf{r}})$
is comprised of a sum of constituent functions, each representative
of a property of the system, with each being the work done in the
corresponding field.

\textbf{(\emph{vi})} Provided the sum of the Lorentz
${\boldsymbol{\cal{L}}} ({\bf{r}})$ and internal magnetic
${\boldsymbol{\cal{I}}}_{m} ({\bf{r}})$ fields is conservative, it
is then possible to define a scalar potential $v_{m} ({\bf{r}})$
representative of all the magnetic effects of the system.  This
potential is the work done in the sum of the fields
${\boldsymbol{\cal{L}}} ({\bf{r}})$ and ${\boldsymbol{\cal{I}}}_{m}
({\bf{r}})$.  This work done is path-independent.

\textbf{(\emph{vii})}  The `Quantal Newtonian' first law also
provides a deeper mathematical understanding of the potential $v
({\bf{r}})$. As the components of the conservative field
${\boldsymbol{\cal{F}}} ({\bf{r}})$ of Eq. (57) are obtained from
quantal sources that are expectation values of Hermitian operators
taken with respect to the wave function $\Psi$, the field
${\boldsymbol{\cal{F}}} ({\bf{r}})$ is a functional of $\Psi$,
\emph{i.e.}  ${\boldsymbol{\cal{F}}} ({\bf{r}}) =
{\boldsymbol{\cal{F}}} [\Psi] ({\bf{r}})$.  This functional is
\emph{exactly known} since the individual component fields are
explicitly defined.  This in turn means that the scalar potential
energy $v ({\bf{r}})$ as defined by Eq. (57) is an \emph{exactly
known} functional of the wave function $\Psi: v ({\bf{r}}) = v
[\Psi] ({\bf{r}})$.  We emphasize that this functional dependence is
valid for \emph{arbitrary} state.  (That the external potential $v
({\bf{r}})$ is a functional of the \emph{ground state} wave function
$\Psi_{g}$ was originally proved by Hohenberg and Kohn \cite{38} for
the case when the only external field present was the electrostatic
binding field ${\boldsymbol{\cal{E}}} ({\bf{r}})$.  The explicit
functional dependence of $v ({\bf{r}})$ on $\Psi_{g}$ was, however,
not given.)

\section{Generalization of the Schr{\"o}dinger-Pauli Equation }

Another consequence of the `Quantal Newtonian' first law is the
generalization of the Schr{\"o}dinger-Pauli equation.  This
generalized form of the equation exhibits its intrinsic
self-consistent nature.   In the previous section, it was shown that
the scalar potential $v ({\bf{r}})$ was a known functional of the
wave function $\Psi$. Substituting the functional $v [\Psi]
({\bf{r}})$  into the Schr{\"o}dinger-Pauli equation Eq. (6), the
equation can then be written as

\begin{eqnarray}
\bigg[ \frac{1}{2} \sum_{k} \big(\hat{\bf{p}}_{k}  + \frac{1}{c}
{\bf{A}} ({\bf{r}}_{k}) \big)^{2} &+& \frac{1}{c} \sum_{k}
{\boldsymbol{\cal{B}}} ({\bf{r}}_{k}) \cdot {\bf{s}}_{k} +
\frac{1}{2} \sideset{}{'}\sum_{k,\ell} \frac{1} {|{\bf{r}}_{k} -
{\bf{r}}_{\ell}|} \nonumber \\
&+& \sum_{k} v [\Psi] ({\bf{r}}_{k}) \bigg] \Psi ({\bf{X}}) = E
[\Psi] \Psi ({\bf{X}}),
\end{eqnarray}
or, on employing Eq. (57), as
\begin{eqnarray}
\bigg[ \frac{1}{2} \sum_{k} \big(\hat{\bf{p}}_{k}  + \frac{1}{c}
{\bf{A}} ({\bf{r}}_{k}) \big)^{2} &+& \frac{1}{c} \sum_{k}
{\boldsymbol{\cal{B}}} ({\bf{r}}_{k}) \cdot {\bf{s}}_{k} +
\frac{1}{2} \sideset{}{'}\sum_{k,\ell} \frac{1} {|{\bf{r}}_{k} -
{\bf{r}}_{\ell}|} \nonumber \\
&+& \sum_{k} \int_{\infty}^{{\bf{r}}_{k}} {\boldsymbol{\cal{F}}}
[\Psi] ({\bf{r}}) \cdot d {\boldsymbol{\ell}} \bigg] \Psi ({\bf{X}})
= E [\Psi] \Psi ({\bf{X}}).
\end{eqnarray}
Thus, the Hamiltonian is a functional of the wave function $\Psi:
\hat{H} = \hat{H} [\Psi]$, and the Schr{\"o}dinger-Pauli equation
can then be written in its generalized form as in Eq. (7).    The
Hamiltonian functional $\hat{H} [\Psi]$ is exactly known.  It is
valid for \emph{arbitrary state}.  It is also \emph{universal} in
that it is applicable to any electronic system defined by this
Hamiltonian.

The generalized form of the Schr{\"o}dinger-Pauli equation makes
evident that its solution $\Psi$ may be obtained self-consistently.
One begins with an appropriate approximate wave function $\Psi$ to
first determine the corresponding quantal sources and fields, and
the potential $v ({\bf{r}})$, and thereby the approximate
Hamiltonian $\hat{H} [\Psi]$. The Schr{\"o}dinger-Pauli equation Eq.
(7 or 59) is then solved to obtain the next approximation to the
wave function $\Psi$ and energy $E$, from which the corresponding
sources and fields and potential $v ({\bf{r}})$ then lead to the
next approximate $\hat{H} [\Psi]$. For this new approximate $\hat{H}
[\Psi]$, the Schr{\"o}dinger-Pauli equation is again solved for the
next approximate wave function $\Psi$ and energy $E$.  And this
procedure is continued till the input wave function $\Psi$ to
$\hat{H} [\Psi]$ is the same $\Psi$ as that generated by this
$\hat{H} [\Psi]$ via solution of the Schr{\"o}dinger-Pauli equation.
Note that the meaning of the functional $v[\Psi]$ is that for each
new $\Psi$, one obtains a new $v[\Psi] ({\bf{r}})$, and therefore
the Hamiltonian functional $\hat{H} [\Psi]$ changes with each new
iterative $\Psi$.   This then allows for the self-consistent
procedure. The understanding that the Schr{\"o}dinger-Pauli equation
is intrinsically self-consistent is new.

In its generalized form, the Schr{\"o}dinger-Pauli equation has
additional attributes, and leads to further insights:

(a)  In traditional quantum mechanics, the potential $v ({\bf{r}})$
is considered as being \emph{extrinsic} to the system of $N$
electrons, and as such is assumed to be a \emph{known} but
\emph{independent} input to the Hamiltonian $\hat{H}$.  In other
words, it does not depend on any other terms of the Hamiltonian
$\hat{H}$.  From the generalized form of the equation, it becomes
evident that the potential $v[\Psi] ({\bf{r}})$ is in fact
\emph{intrinsic} to the physical system being related to it via the
internal field components (see Eq. (57)).  It is thereby
(self-consistently) \emph{dependent} on all the properties of the
system via the other operators of the Hamiltonian $\hat{H}$.

(b) On achieving self-consistency, the wave function $\Psi
({\bf{X}})$, the eigen energy $E$, and the potential $v[\Psi]
({\bf{r}})$ or equivalently the Hamiltonian $\hat{H} [\Psi]$, are
determined. This is of particular significance in those cases for
which the potential $v ({\bf{r}})$ may be \emph{unknown}.  Due to
the advances in semiconductor technology, it has been possible to
create 2-dimensional `artificial atoms' or quantum dots.  When such
quantum dots were initially developed, the form of the binding
potential of the electrons was not known. Later, via experimentation
and theoretical work at the Hartree level \cite{13,14,54}, it was
determined that the potential was harmonic. This is now accepted to
be the case.  Had the generalized form of the Schr{\"o}dinger-Pauli
equation existed at that time, the fact of a harmonic binding
potential $v ({\bf{r}})$ for quantum dots could have been arrived at
via its self-consistent solution.   In the future, when new
electronic devices are created, the corresponding binding potential
could thus be obtained.

(c)  The self-consistent procedure could also be employed to
determine the wave function $\Psi ({\bf{X}})$ and energy $E$ even
for the common case when the potential $v ({\bf{r}})$ is known.
Starting with an accurate approximate wave function $\Psi$, the
corresponding approximate $v[\Psi] ({\bf{r}})$ could be determined.
Of course, this would not correspond to the known $v ({\bf{r}})$
function.  But the solution of the resulting Schr{\"o}dinger-Pauli
equation with this approximate $v[\Psi] ({\bf{r}})$ would be an
improvement to the original wave function.  Continuing with the
self-consistency procedure would lead to the exact $\{ \Psi
({\bf{X}}), E \}$.  On achieving self-consistency, the known
function $v ({\bf{r}})$ would be reproduced.

(d)  It is worth comparing the self-consistent procedure of the
generalized Schr{\"o}dinger-Pauli  equation with that of the
variational principle for the \emph{ground} state energy \cite{55}.
Starting with an approximate parameterized wave function correct to
$O (\delta)$, the variational principle leads to a rigorous upper
bound to the energy that is correct to $O (\delta^{2})$.  However,
all other properties of the system are obtained correct to only $O
(\delta)$. Thus, the variationally obtained wave function is
accurate only in the region of space contributing to the energy.  On
the other hand, the self-consistently obtained solution is accurate
throughout space, and hence all properties are accurate to the same
degree of numerical accuracy as required.   For \emph{excited}
states, the application of the variational principle requires that
the approximate wave function be orthogonal to the exact ground
state wave function.  The generalized Schr{\"o}dinger-Pauli equation
is valid for both ground and excited states.  The corresponding
excited state wave function obtained self-consistently will
automatically be orthogonal to the ground and other states of the
system.

(e)  In the Schr{\"o}dinger theory \cite{30} of electrons in the
presence of electrostatic ${\boldsymbol{\cal{E}}} ({\bf{r}})$ and
magnetostatic ${\boldsymbol{\cal{B}}} ({\bf{r}})$ fields, one hews
to the philosophy that electromagnetic interactions occur by the
substitution $\hat{\bf{p}} \rightarrow  \hat{\bf{p}} + (e/c)
{\bf{A}}$. Thus, it is the vector potential ${\bf{A}} ({\bf{r}})$
and not the magnetic field ${\boldsymbol{\cal{B}}} ({\bf{r}})$ that
appears in the corresponding Schr{\"o}dinger equation. This
fundamental difference between classical and quantum physics then
explains, for example, the Aharonov-Bohm effect \cite{56}.  The
magnetic field ${\boldsymbol{\cal{B}}} ({\bf{r}})$ appears in the
Schr{\"o}dinger equation only after a choice of gauge for the vector
potential ${\bf{A}} ({\bf{r}})$.  In the Schr{\"o}dinger-Pauli
equation (Eq. (6)), the magnetic field ${\boldsymbol{\cal{B}}}
({\bf{r}})$ appears explicitly as a consequence of the use of the
Feynman kinetic energy operator $\hat{T}_{F}$: It is the term
corresponding to the interaction between the magnetic field and the
spin angular momentum operator. The generalized form of the
Schr{\"o}dinger-Pauli equation further shows that the magnetic field
${\boldsymbol{\cal{B}}} ({\bf{r}})$ also appears in the term
involving $v[\Psi] ({\bf{r}})$ via the conservative field
${\boldsymbol{\cal{F}}} ({\bf{r}})$ which includes the Lorentz field
${\boldsymbol{\cal{L}}} ({\bf{r}})$ (see Eqs. 57-59).

\section{Quantal Density Functional Theory}

In this section the system of electrons described by the
Schr{\"o}dinger-Pauli Hamiltonian of Eq. (5) is mapped via quantal
density functional theory (QDFT) \cite{2,3} to one of noninteracting
fermions possessing the same electronic $\rho ({\bf{r}})$ and
physical current ${\bf{j}} ({\bf{r}})$ density.  The additional
constraints on the model system are that it also possesses the same
number $N$ of fermions, and the same total orbital ${\bf{L}}$ and
spin ${\bf{S}}$ angular momentum. It is assumed that the model
fermions are subject to the same electrostatic
${\boldsymbol{\cal{E}}} ({\bf{r}}) = - {\boldsymbol{\nabla}} v
({\bf{r}})/e$ and magnetostatic ${\boldsymbol{\cal{B}}} ({\bf{r}}) =
{\boldsymbol{\nabla}} \times {\bf{A}} ({\bf{r}})$ fields as the
electrons of the interacting system. It is further assumed that such
a model system can exist.

The key to the mapping from the interacting to the noninteracting
fermion model system is to determine the \emph{local}
electron-interaction potential $v_\mathrm{ee} ({\bf{r}})$ in which
the many-body effects are incorporated. This potential then
generates the single-particle orbitals $\phi_{i} ({\bf{x}})$ of the
Slater determinant $\Phi \{ \phi_{i} ({\bf{x}}) \}$ that lead to the
same electronic and physical current density.  For the QDFT model
system with the constraints as described above, the only electron
correlations that must be \emph{explicitly} accounted for in
$v_\mathrm{ee} ({\bf{r}})$ are those due to the Pauli exclusion
principle and Coulomb repulsion, and Correlation-Kinetic effects
\cite{57}.

It is reiterated that a principal purpose of the mapping to the
model system is to determine the Correlation-Kinetic energy $T_{c}$,
and to separate the Pauli-Coulomb $E_{xc}$ (quantum-mechanical
exchange-correlation) energy into its Pauli $E_{x}$ (exchange) and
Coulomb $E_{c}$ (correlation) energy components.

Consider a system of $N$ noninteracting fermions possessing the same
potential energies $\{ v, {\bf{A}} \}$ as that of the interacting
electrons.  The Schr{\"o}dinger-Pauli Hamiltonian $\hat{H}_{s}$ of
the model fermions (the $S$ system) is (see also Eq. (16))
\begin{equation}
\hat{H}_{s} = \hat{T} + \frac{1}{c} \int \hat{\bf{j}}_{s} ({\bf{r}})
\cdot {\bf{A}} ({\bf{r}}) d {\bf{r}} - \frac{1} {2c^{2}} \int
\hat{\rho} ({\bf{r}}) A^{2} ({\bf{r}}) d {\bf{r}} + \hat{V}_{s},
\end{equation}
where the local potential operator $\hat{V}_{s}$ is
\begin{equation}
\hat{V}_{s} = \sum_{k} v_{s} ({\bf{r}}_{k}) = \sum_{k} \bigg[ v
({\bf{r}}_{k}) + v_\mathrm{ee} ({\bf{r}}_{k}) \bigg],
\end{equation}
and $v_\mathrm{ee} ({\bf{r}})$ the \emph{local} electron-interaction
potential in which all the many-body effects are incorporated.  As
the configuration of the model fermions is as yet unspecified, the
$S$ system current density operator $\hat{\bf{j}}_{s} ({\bf{r}})$ is
\begin{equation}
\hat{\bf{j}}_{s} ({\bf{r}}) = \hat{\bf{j}}_{p} ({\bf{r}}) +
\hat{\bf{j}}_{d} ({\bf{r}}) + \hat{\bf{j}}_{m,s} ({\bf{r}}),
\end{equation}
with the paramagnetic $\hat{\bf{j}}_{p} ({\bf{r}})$ and diamagnetic
$\hat{\bf{j}}_{d} ({\bf{r}})$ current density operators defined as
in Eqs. (10) and (11).  The magnetization current density operator
$\hat{\bf{j}}_{m,s} ({\bf{r}}) = -c {\boldsymbol{\nabla}} \times
\hat{\bf{m}}_{s} ({\bf{r}})$ with the magnetization density operator
$\hat{\bf{m}}_{s} ({\bf{r}}) = (-\frac{1}{c}) \sum_{k}
{\bf{s}}_{k,s} \delta ({\bf{r}}_{k} - {\bf{r}})$, and
${\bf{s}}_{k,s}$ the spin vector of the $k$-th model fermion.

For \emph{arbitrary state} of the interacting system, the mapping to
the model system is to be such that it possesses the same basic
variables $\{ \rho ({\bf{r}}), {\bf{j}} ({\bf{r}}) \}$ and satisfies
the same constraints on $N, {\bf{L}}$, and ${\bf{S}}$.  With the
orbital angular momentum ${\bf{L}}$ being the same, the equivalence
of the spin ${\bf{S}}$ angular momentum requires that
${\bf{s}}_{k,s} = {\bf{s}}_{k}$.  (This means that the configuration
of the model fermions is either the same as that of the interacting
electrons, or a different configuration but one possessing the same
${\bf{L}}$ and ${\bf{S}}$. Thus, for example, it is possible to map
an interacting two-electron system in an excited singlet state  to a
$S$ system in its ground state.)  The equivalence of the spin
vectors implies that $\hat{\bf{m}}_{s} ({\bf{r}}) = {\bf{m}}
({\bf{r}})$ so that $\hat{\bf{j}}_{m,s} ({\bf{r}}) =
\hat{\bf{j}}_{m} ({\bf{r}})$.  It follows that the operator
$\hat{\bf{j}}_{s} ({\bf{r}}) = \hat{\bf{j}} ({\bf{r}})$.  Hence, the
$S$ system Hamiltonian $\hat{H}_{s}$ of Eq. (60) may be written as
\begin{equation}
\hat{H}_{s} = \hat{T} + \frac{1}{c} \int \hat{\bf{j}} ({\bf{r}})
\cdot {\bf{A}} ({\bf{r}}) d {\bf{r}} - \frac{1}{2c^{2}} \int
\hat{\rho} ({\bf{r}}) A^{2} ({\bf{r}}) + \hat{V}_{s}.
\end{equation}
The corresponding local effective potential differential equation
for the orbitals $\phi_{i} ({\bf{x}})$ of the Slater determinant
wave function $\Phi \{ \phi_{k} \}$ of the model fermions (assuming
additionally that $c=1$) is
\begin{equation}
\bigg[ \frac{1}{2} (\hat{\bf{p}} + {\bf{A}} ({\bf{r}}) )^{2} +
{\bf{B}} \cdot {\bf{s}} + v ({\bf{r}}) + v_\mathrm{ee} ({\bf{r}})
\bigg] \phi_{k} ({\bf{x}}) = \epsilon_{k} \phi_{k} ({\bf{x}}) ~;~
k=1, \ldots, N.
\end{equation}
The $S$ system properties of the density $\rho_{s} ({\bf{r}})$,
Dirac density matrix $\gamma_{s} ({\bf{r r}}')$, pair-correlation
density $g_{s} ({\bf{r r}}')$, and the current density ${\bf{j}}_{s}
({\bf{r}})$ are then, respectively, the expectation values of the
corresponding Hermitian operators taken with respect to the Slater
determinant wave function $\Phi \{ \phi_{i} \}$.  Thus $\rho_{s}
({\bf{r}}) = \langle \Phi \{ \phi_{k} \} \hat{\rho} |({\bf{r}})|
\Phi \{ \phi_{k} \} \rangle = \sum_{\sigma, k} \phi^{\star}_{k}
({\bf{r}} \sigma) \phi_{k} ({\bf{r}} \sigma)$; $\gamma_{s} ({\bf{r
r}}') = \langle \Phi \{ \phi_{k} \} |\hat{\gamma} ({\bf{r r}}')|
\Phi \{ \phi_{k} \} \rangle = \sum_{\sigma,k} \phi^{\star}_{k}
({\bf{r}} \sigma) \phi_{k} ({\bf{r}}' \sigma)$; $g_{s} ({\bf{r r}}')
= \frac{1} {\rho_{s} ({\bf{r}})} \langle \Phi \{ \phi_{k} \}
|\hat{P} ({\bf{r r}}')| \Phi \{ \phi_{k} \} \rangle$; ${\bf{j}}_{s}
({\bf{r}}) = \langle \Phi \{ \phi_{k} \}|\hat{\bf{j}} ({\bf{r}})
|\Phi \{ \phi_{k} \} \rangle$.  Note that $g_{s} ({\bf{r r}}') =
\rho ({\bf{r}}') + \rho_{x} ({\bf{r r}}')$, where the Fermi hole is
defined as $\rho_{x} ({\bf{r r}}') = - |\gamma_{s} ({\bf{r
r}}')|^{2}/2 \rho_{s} ({\bf{r}})$.

With the requirement that the $S$ system density $\rho_{s}
({\bf{r}})$ and current density ${\bf{j}}_{s} ({\bf{r}})$ are the
same as $\{ \rho ({\bf{r}}), {\bf{j}} ({\bf{r}}) \}$ of the
interacting system, the `Quantal Newtonian' first law as satisfied
by each model fermion is then
\begin{equation}
{\boldsymbol{\cal{F}}}^\mathrm{ext} ({\bf{r}}) +
{\boldsymbol{\cal{F}}}^\mathrm{int}_{s} ({\bf{r}}) = 0.
\end{equation}
The law is derived (see Appendix) employing the continunity
condition ${\boldsymbol{\nabla}} \cdot {\bf{j}} ({\bf{r}}) = 0$.  As
the potentials $\{ v, {\bf{A}} \}$ and the densities $\{ \rho
({\bf{r}}), {\bf{j}} ({\bf{r}}) \}$ of the $S$ system are the same
as those of the interacting system, the external field
${\boldsymbol{\cal{F}}}^\mathrm{ext} ({\bf{r}})$ experienced by the
model fermions is the same as for the electrons (see Eq. (17)).  The
internal field ${\boldsymbol{\cal{F}}}^\mathrm{int}_{s} ({\bf{r}})$
of these fermions is obtained as
\begin{equation}
{\boldsymbol{\cal{F}}}^\mathrm{int}_{s} ({\bf{r}}) = -
{\boldsymbol{\nabla}} v_\mathrm{ee} ({\bf{r}}) -
{\boldsymbol{\cal{Z}}}_{s} ({\bf{r}}) - {\boldsymbol{\cal{D}}}
({\bf{r}}) - {\boldsymbol{\cal{I}}}_{m} ({\bf{r}}),
\end{equation}
where ${\boldsymbol{\cal{Z}}}_{s} ({\bf{r}}), {\boldsymbol{\cal{D}}}
({\bf{r}}), {\boldsymbol{\cal{I}}}_{m} ({\bf{r}})$ are the
corresponding kinetic, differential density, and internal magnetic
fields.  The $S$ system kinetic field ${\boldsymbol{\cal{Z}}}_{s}
({\bf{r}})$ is defined in a manner similar to the kinetic field
${\boldsymbol{\cal{Z}}} ({\bf{r}})$ of the interacting system (see
Eq. (30)):
\begin{equation}
{\boldsymbol{\cal{Z}}}_{s} ({\bf{r}}) = \frac{{\bf{z}}_{s}
({\bf{r}})} {\rho ({\bf{r}})},
\end{equation}
where the kinetic `force' ${\bf{z}}_{s} ({\bf{r}})$ is obtained from
its \emph{nonlocal} quantal source, the Dirac density matrix
$\gamma_{s} ({\bf{r r}}')$ (defined earlier) as
\begin{equation}
z_{s, \alpha} ({\bf{r}}) = 2 \sum_{\beta} \nabla_{\beta} t_{s,
\alpha \beta} ({\bf{r}}; \gamma_{s}),
\end{equation}
where the second rank tensor $t_{s, \alpha \beta} ({\bf{r}};
\gamma_{s})$ is
\begin{equation}
t_{s, \alpha \beta} ({\bf{r}}; \gamma_{s}) = \frac{1}{4} \bigg[
\frac{\partial^{2}} {\partial r'_{\alpha} \partial r''_{\beta}} +
\frac{\partial^{2}} {\partial r'_{\beta} \partial r''_{\alpha}}
\bigg] \gamma_{s} ({\bf{r'}} {\bf{r}}'') \bigg|_{{\bf{r}}' =
{\bf{r}}'' = r}.
\end{equation}
The fields ${\boldsymbol{\cal{D}}} ({\bf{r}})$ and
${\boldsymbol{\cal{I}}}_{m} ({\bf{r}})$ are defined as for the
interacting system.  As the densities $\{ \rho ({\bf{r}}), {\bf{j}}
({\bf{r}}) \}$ of the interacting and $S$ systems are the same, so
are these corresponding fields (see Eqs. (37) and (39)).

Equating the `Quantal Newtonian' first laws for the interacting and
model systems (Eqs. (42) and (65)) then leads to the definition of
the \emph{local} electron-interaction potential $v_\mathrm{ee}
({\bf{r}})$ of the $S$ system differential equation (Eq. (64)).
\emph{The potential $v_\mathrm{ee} ({\bf{r}})$ is the work done to
move the model fermion from some reference point at infinity to its
position at ${\bf{r}}$ in the force of a conservative effective
field ${\boldsymbol{\cal{F}}}^\mathrm{eff} ({\bf{r}})$}:
\begin{equation}
v_\mathrm{ee} ({\bf{r}}) = - \int_{\infty}^{{\bf{r}}}
{\boldsymbol{\cal{F}}}^\mathrm{eff} ({\bf{r}}') \cdot d
{\boldsymbol{\ell}}',
\end{equation}
where
\begin{equation}
{\boldsymbol{\cal{F}}}^\mathrm{eff} ({\bf{r}}) =
{\boldsymbol{\cal{E}}}_\mathrm{ee} ({\bf{r}}) +
{\boldsymbol{\cal{Z}}}_{t_{c}} ({\bf{r}}),
\end{equation}
with the electron-interaction field
${\boldsymbol{\cal{E}}}_\mathrm{ee} ({\bf{r}})$ given by Eq. (23),
and the correlation-kinetic field ${\boldsymbol{\cal{Z}}}_{t_{c}}
({\bf{r}})$ defined as
\begin{equation}
{\boldsymbol{\cal{Z}}}_{t_{c}} ({\bf{r}}) =
{\boldsymbol{\cal{Z}}}_{s} ({\bf{r}}) - {\boldsymbol{\cal{Z}}}
({\bf{r}}).
\end{equation}
Since the ${\boldsymbol{\nabla}} \times
{\boldsymbol{\cal{F}}}^\mathrm{eff} ({\bf{r}})$ vanishes, the
potential $v_\mathrm{ee} ({\bf{r}})$ is \emph{path-independent}.
Further, in the self-consistent determination of $v_\mathrm{ee}
({\bf{r}})$, it follows from Eq. (71) that the only correlations
that must be accounted for are those due to the Pauli exclusion
principle, Coulomb repulsion, and correlation-kinetic effects.

The total energy $E$ of the interacting system can be expressed in
terms of the $S$ system properties.  Splitting the kinetic energy
$T$ into its noninteracting $T_{s}$ and correlation-kinetic $T_{c}$
components, the energy $E = \langle \Psi ({\bf{X}}) | \hat{H}| \Psi
({\bf{X}}) \rangle$ (assuming $c=1$) may be written as (see Eq.
(16))
\begin{equation}
E = T_{s} + \int \rho ({\bf{r}}) v ({\bf{r}}) d {\bf{r}} + \int
{\bf{j}} ({\bf{r}}) \cdot {\bf{A}} ({\bf{r}}) d {\bf{r}} -
\frac{1}{2} \int \rho ({\bf{r}}) A^{2} ({\bf{r}}) d {\bf{r}} +
E_\mathrm{ee} + T_{c}.
\end{equation}
By multiplying the $S$ system differential equation Eq. (64) by
$\phi^{\star}_{k} ({\bf{x}})$, summing over all the model fermions,
and integrating over all space, the noninteracting kinetic energy
$T_{s}$ is obtained as
\begin{equation}
T_{s} = \sum_{k} \epsilon_{k} - \int \rho ({\bf{r}}) v ({\bf{r}}) d
{\bf{r}} - \int \rho ({\bf{r}}) v_\mathrm{ee} ({\bf{r}}) d {\bf{r}}
- \int {\bf{j}} ({\bf{r}}) \cdot {\bf{A}} ({\bf{r}}) d {\bf{r}} +
\frac{1}{2} \int \rho ({\bf{r}}) A^{2} ({\bf{r}}) d {\bf{r}}.
\end{equation}
In substituting Eq. (74) into Eq. (73) the expression for $E$
becomes
\begin{equation}
E = \sum_{k} \epsilon_{k} - \int \rho ({\bf{r}}) v_\mathrm{ee}
({\bf{r}}) d {\bf{r}} + E_\mathrm{ee} + T_{c},
\end{equation}
where
\begin{equation}
T_{c} = \frac{1}{2} \int \rho ({\bf{r}}) {\bf{r}} \cdot
{\boldsymbol{\cal{Z}}}_{t_{c}} ({\bf{r}}) d {\bf{r}}.
\end{equation}
Thus, the correlation-kinetic energy $T_{c}$ -- the contribution of
electron correlations to the kinetic energy -- is explicitly
defined.  This is a property of the electronic system not obtainable
solely by solution of the Schr{\"o}dinger-Pauli equation.

Finally, via the mapping to the model $S$ system, it is possible to
split the Pauli-Coulomb energy $E_{xc}$ of Eq. (47) into its Pauli
$E_{x}$ and Coulomb $E_{c}$ components.  Defining the Coulomb hole
$\rho_{c} ({\bf{r r}}')$ as the difference between the Fermi-Coulomb
$\rho_{xc} ({\bf{r r}}')$ and Fermi $\rho_{x} ({\bf{r r}}')$ hole
charges where $\rho_{x} ({\bf{r r}}')$ is determinmed from
$\gamma_{s} ({\bf{r r}}')$ as mentioned earlier:  $\rho_{c} ({\bf{r
r}}') = \rho_{xc} ({\bf{r r}}') - \rho_{x} ({\bf{r r}}')$, the
Pauli-Coulomb field ${\boldsymbol{\cal{E}}}_{xc} ({\bf{r}})$ of Eq.
(29) may be written as a sum of its Pauli
${\boldsymbol{\cal{E}}}_{x} ({\bf{r}})$ and Coulomb
${\boldsymbol{\cal{E}}}_{c} ({\bf{r}})$ components:
\begin{equation}
{\boldsymbol{\cal{E}}}_{xc} ({\bf{r}}) = {\boldsymbol{\cal{E}}}_{x}
({\bf{r}}) + {\boldsymbol{\cal{E}}}_{c} ({\bf{r}}),
\end{equation}
where
\begin{equation}
{\boldsymbol{\cal{E}}}_{x} ({\bf{r}}) = \int \frac{\rho_{x} ({\bf{r
r}}') ({\bf{r}} - {\bf{r}}')} {| {\bf{r}} - {\bf{r}}'|^{3}} d
{\bf{r}}',
\end{equation}
and
\begin{equation}
{\boldsymbol{\cal{E}}}_{c} ({\bf{r}}) = \int \frac{\rho_{c} ({\bf{r
r}}') ({\bf{r}} - {\bf{r}}')} {| {\bf{r}} - {\bf{r}}'|^{3}} d
{\bf{r}}'.
\end{equation}
The corresponding Pauli $E_{x}$ and Coulomb $E_{c}$ energies are
then, respectively,
\begin{equation}
E_{x} = \int \rho ({\bf{r}}) {\bf{r}} \cdot
{\boldsymbol{\cal{E}}}_{x} ({\bf{r}}) d {\bf{r}},
\end{equation}
and
\begin{equation}
E_{c} = \int \rho ({\bf{r}}) {\bf{r}} \cdot
{\boldsymbol{\cal{E}}}_{c} ({\bf{r}}) d {\bf{r}}.
\end{equation}
The total energy $E$ of the interacting system may thus be written
in terms of its components as (see Eq. (56))
\begin{equation}
E = T_{s} + E_{H} + E_{x} + E_{c} + T_{c} + E_{es} + E_\mathrm{mag}.
\end{equation}
(Note that $T_{s}$ may be determined either via the field
${\boldsymbol{\cal{Z}}}_{s} ({\bf{r}})$ through its integral virial
expression, or as the expectation value $\langle \Phi \{ \phi_{k} \}
|\hat{T}| \Phi \{ \phi_{k} \} \rangle$).  It is emphasized that the
components of the energy $E$ are properties of the \emph{same}
physical system.

The fact that the energy $E$ can be so expressed in terms of these
individual components shows the significance of the QDFT mapping to
the model system.  As such the mapping constitutes an essential
complement to Schr{\"o}dinger-Pauli theory.  The mapping to the
model system possessing the same basic variables of the density and
physical current density provides a deeper insight into the physical
system.

The second principal reason for the development of a local effective
potential theory such as the QDFT described above, or Kohn-Sham
density functional theory, or the Optimized Potential Method, is the
easier numerical solution of the corresponding single-particle
differential or integro-differential equation.  These theories, of
course, lead to properties of the interacting system.  The QDFT
differential equation Eq. (64) is designed to deliver the
interacting system density $\rho ({\bf{r}})$, and the current
density ${\bf{j}} ({\bf{r}})$. From these properties may be
determined all single-particle operator expectation values, the
Lorentz `force' and field, and the internal magnetic 'force' and
field.  The highest occupied eigenvalue of the differential equation
is the negative of the ionization potential. The total energy $E$ is
in turn determined via Eq. (75).  In the application of local
effective potential theories, approximations must of course be made.
For a description of approximation methods within QDFT, such as
those of many-body perturbation theory, the M\o ller-Plesset
perturbation theory, or approximations based on electron
correlations, the reader is referred to Ref. [3].

\section{Summary of New Understandings and Future Work}

In this paper, the Schr{\"o}dinger-Pauli theory of electrons in a
static electromagnetic field is described from a new perspective,
one that leads to further insights into the quantum-mechanical
description of the physical system, and thereby into the quantum
mechanics of electronic structure.  The perspective is that of the
\emph{individual electron} via its equation of motion, the `Quantal
Newtonian' first law.  The law is in terms of `classical' fields
that pervade all space.  The fields arise from quantal sources that
are expectation values of Hermitian operators taken with respect to
the system wave function $\Psi$.  Hence, the perspective hews to the
Copenhagen probabilistic interpretation of the wave function via
these quantal sources.  The fields obey the equations of classical
physics, and are therefore determinate.  This determinism is in the
same sense as those of the classical electrostatic field between two
charges or magnetostatic field between two magnetic poles.  In this
context, the new perspective is simultaneously  probabilistic and
deterministic.

As is the case for classical particles in an external field and
interacting via Newton's third law forces, the `Quantal Newtonian'
first law is comprised of the vanishing sum of the external
${\boldsymbol{\cal{F}}}^\mathrm{ext} ({\bf{r}})$ and internal
${\boldsymbol{\cal{F}}}^\mathrm{int} ({\bf{r}})$ fields experienced
by each electron.  The external field
${\boldsymbol{\cal{F}}}^\mathrm{ext} ({\bf{r}})$ is comprised of the
sum of the electrostatic ${\boldsymbol{\cal{E}}} ({\bf{r}})$ and
Lorentz ${\boldsymbol{\cal{L}}} ({\bf{r}})$ fields.  The latter
arises from its quantal source, the physical current density
${\bf{j}} ({\bf{r}})$ via the Lorentz `force'.  In the
Schr{\"o}dinger-Pauli differential equation, the presence of the
scalar potential $v ({\bf{r}})$ implies the existence of the
electrostatic field ${\boldsymbol{\cal{E}}} ({\bf{r}})$.  However,
the fact that each electron experiences a Lorentz `force' or field,
though implicitly understood to be the case, is not explicitly
represented by a term in the Schr{\"o}dinger-Pauli differential
equation or in the Schr{\"o}dinger equation for electrons in the
presence of a magnetic field.   Such a term appears
\emph{explicitly} in the `Quantal Newtonian' first law. Hence, the
`Quantal Newtonian' first law makes explicit our understanding that
in the presence of a magnetic field, each electron experiences a
Lorentz field ${\boldsymbol{\cal{L}}} ({\bf{r}})$ and `force'.

The `Quantal Newtonian' first law also informs that each electron
experiences an internal field ${\boldsymbol{\cal{F}}}^\mathrm{int}
({\bf{r}})$ comprised of a sum of fields each representative of a
property of the system.  Thus, the electron-interaction
${\boldsymbol{\cal{E}}}_\mathrm{ee} ({\bf{r}})$, differential
density ${\boldsymbol{\cal{D}}} ({\bf{r}})$, and kinetic
${\boldsymbol{\cal{Z}}} ({\bf{r}})$ fields are representative of the
electron-correlations due to the Pauli exclusion principle and
Coulomb repulsion, the electronic density, and kinetic effects,
respectively.  There is also a magnetic field component
${\boldsymbol{\cal{I}}}_{m} ({\bf{r}})$  to the internal field. The
fact that each electron is furthermore subject to these
property-related fields is also not evident from
Schr{\"o}dinger-Pauli theory as presently understood.

The external magnetic field ${\boldsymbol{\cal{B}}} ({\bf{r}})$ thus
gives rise in quantum mechanics to both the Lorentz field
${\boldsymbol{\cal{L}}} ({\bf{r}})$ as well as an internal magnetic
field ${\boldsymbol{\cal{I}}}_{m} ({\bf{r}})$ as experienced by each
electron. Interestingly, if the sum of these fields is conservative,
then it is possible to describe the contributions of the magnetic
field to the quantum system by a scalar (path-independent) magnetic
potential $v_{m} ({\bf{r}})$ similar to the external scalar
electrostatic potential $v ({\bf{r}})$.

Another new understanding arrived at via the `Quantal Newtonian'
first law is that the Schr{\"o}dinger-Pauli Hamiltonian $\hat{H}$ is
a functional of the wave function $\Psi$, i.e. $\hat{H} = \hat{H}
[\Psi]$. This functional is \emph{exactly known} and
\emph{universal}. With the Schr{\"o}dinger-Pauli equation now
written as $\hat{H} [\Psi] \Psi = E[\Psi] \Psi$, it becomes evident
that the equation is intrinsically self-consistent. This then allows
for the self-consistent determination of the wave function $\Psi$
and eigen energy $E$.  It also allows for the determination of the
external scalar potential $v ({\bf{r}})$ for new physical systems
that may be created in the future for which the binding potential is
unknown. The traditional approach to Schr{\"o}dinger-Pauli theory,
(with the binding potential $v ({\bf{r}})$ known), has been the
direct solution of the differential equation for the wave function
$\Psi$. The issue of whether the Schr{\"o}dinger-Pauli differential
equation was a self-consistent one did not arise.  On the other
hand, all single-particle formalisms such as Hartree, Hartree-Fock,
and local effective potential theories, which are derivatives of the
Schr{\"o}dinger-Pauli theory, are intrinsically self-consistent.  We
now understand that the fundamental equation on which these theories
are founded -- the Schr{\"o}dinger-Pauli equation -- is itself
self-consistent.

In order to obtain additional properties, the (interacting) physical
system is mapped via the quantal source-field perspective of QDFT to
one of noninteracting  fermions possessing the same basic variables
of the electronic density $\rho ({\bf{r}})$ and physical current
density ${\bf{j}} ({\bf{r}})$, and the same electron number $N$,
orbital ${\bf{L}}$ and spin ${\bf{S}}$ angular momentum. The
additional properties obtained thereby are the correlation
contribution to the kinetic energy - the correlation-kinetic energy
$T_{c}$; the contribution of the electron correlations due to the
Pauli exclusion principle to the energy -- the Pauli energy $E_{x}$;
the correlation contribution to the energy beyond the Hartree energy
$E_{H}$ due to Coulomb repulsion -- the Coulomb energy $E_{c}$; the
ionization potential or electron affinity. In this manner, the QDFT
mapping constitutes an essential complement to Schr{\"o}dinger-Pauli
theory. The model system can also be thought of as being an
independent local effective potential theory in which each model
fermion experiences the same effective field, and therefore the same
effective potential.  This allows for an easier numerical solution
of the corresponding differential equation.

A generalization of the stationary-state Schr{\"o}dinger-Pauli
theory as described above would be the extension to the temporal
case.  Hence, in addition to the external binding electrostatic
field ${\boldsymbol{\cal{E}}} ({\bf{r}}) = - {\boldsymbol{\nabla}} v
({\bf{r}})/e$, the electrons would be subject to a time-dependent
electromagnetic field: ${\bf{E}} ({\bf{y}}) = -
{\boldsymbol{\nabla}} \phi ({\bf{y}}) - (1/c) \partial {\bf{A}}
({\bf{y}}) / \partial t$ ; ${\boldsymbol{\cal{B}}} ({\bf{y}}) =
{\boldsymbol{\nabla}} \times  {\bf{A}} ({\bf{y}})$, with ${\bf{y}} =
({\bf{r}}, t)$. This would then lead to the time-dependent equation
of motion for each electron or equivalently the `Quantal Newtonian'
second law. The law would then give rise to further insights into
time-dependent Schr{\"o}dinger-Pauli theory as in the present work.
One could go beyond the Born-Oppenheimer approximation by assuming a
time-dependent binding potential ${\boldsymbol{\cal{E}}} ({\bf{y}})
= - {\boldsymbol{\nabla}} v ({\bf{y}})/e$.

The stationary-state Schr{\"o}dinger-Pauli equation can be derived
as the non-relativistic limit of the time-independent Dirac
equation. For a particle of charge $q$ and mass $m$ in a static
electromagnetic field defined by the potentials $\{ v, {\bf{A}} \}$,
the Dirac equation for the 2-component spinors $\psi ({\bf{r}})$ and
$\eta ({\bf{r}})$ which make up the four-component spinor $\chi
({\bf{r}})$, the solution to the Dirac equation, reduces to the
coupled equations :
\begin{equation}
E \psi ({\bf{r}}) = c \hat{\bf{p}}_\mathrm{phys} \cdot
{\boldsymbol{\sigma}} \eta ({\bf{r}}) +  (q v + mc^{2}) \psi
({\bf{r}}),
\end{equation}
\begin{equation}
E \eta ({\bf{r}}) = c \hat{\bf{p}}_\mathrm{phys} \cdot
{\boldsymbol{\sigma}} \psi ({\bf{r}}) +  (q v - mc^{2}) \eta
({\bf{r}}).
\end{equation}
In the non-relativistic limit, the small component $\eta ({\bf{r}})$
can be written in terms of the large component $\psi ({\bf{r}})$.
Substituting this expression for $\eta ({\bf{r}})$ into Eq. (83)
then leads to the Schr{\"o}dinger-Pauli equation for the particle.
It is first proposed to further generalize the ideas presented in
this paper to the above Dirac equation, and then to extend them to
the many particle case.

\section{Acknowledgment}
I thank Prof. Marlina Slamet for her critical reading of the
manuscript.

\newpage
\appendix
\section{Derivation of the `Quantal Newtonian' First Law for
Schr{\"o}dinger-Pauli Theory}

Consider first a system of $N$ \emph{spin-less electrons} in an
external electrostatic ${\boldsymbol{\cal{E}}} ({\bf{r}}) = -
{\boldsymbol{\nabla}} v ({\bf{r}})/e$ and magnetostatic
${\boldsymbol{\cal{B}}} ({\bf{r}}) = {\boldsymbol{\nabla}} \times
{\bf{A}} ({\bf{r}})$ field.  The Schr{\"o}dinger-Pauli theory
equation for the system is
\begin{equation}
\hat{H}_\mathrm{spin-less} \Psi = E \Psi,
\end{equation}
where the Hamiltonian $\hat{H}_\mathrm{spin-less}$ (charge of
electron $-e, |e| = \hbar = m =1$)
\begin{equation}
\hat{H}_\mathrm{spin-less} = \hat{T}_{A} + \hat{W} + \hat{V},
\end{equation}
with
\begin{equation}
\hat{T}_{A} = \frac{1}{2} \sum_{k} \bigg( \hat{\bf{p}}_{k} +
\frac{1}{c} {\bf{A}} ({\bf{r}}_{k}) \bigg)^{2} ~ ;~ \hat{W} =
\frac{1}{2} \sideset{}{'}\sum_{k,\ell} \frac{1}{| {\bf{r}}_{k} -
{\bf{r}}_{\ell}|} ~ ; ~ \hat{V} = \sum_{k} v ({\bf{r}}_{k}),
\end{equation}
the physical kinetic, electron-interaction, and scalar potential
operators, respectively.

A method \cite{2,18,58,59,60} for deriving the `Quantal Newtonian'
first law in general is to write the wave function as $\Psi =
\Psi^{R} + i \Psi^{I}$, where $\Psi^{R}$ and $\Psi^{I}$ are the real
and imaginary parts, substitute it into the corresponding
differential equation, perform the various derivatives, employ the
equation of continunity, and after considerable algebra \cite{18},
arrive at the law.  The law for the spin-less electron is the
vanishing of the sum of an external
${\boldsymbol{\cal{F}}}^\mathrm{ext} ({\bf{r}})$ and internal
${\boldsymbol{\cal{F}}}^\mathrm{int} ({\bf{r}})$ fields.
\begin{equation}
{\boldsymbol{\cal{F}}}^\mathrm{ext} ({\bf{r}}) +
{\boldsymbol{\cal{F}}}^\mathrm{int} ({\bf{r}}) = 0.
\end{equation}
where
\begin{eqnarray}
{\boldsymbol{\cal{F}}}^\mathrm{ext} ({\bf{r}}) &=&
{\boldsymbol{\cal{E}}} ({\bf{r}}) - {\boldsymbol{\cal{L}}}
({\bf{r}}), \\
{\boldsymbol{\cal{F}}}^\mathrm{int} ({\bf{r}}) &=&
{\boldsymbol{\cal{E}}}_\mathrm{ee} ({\bf{r}}) -
{\boldsymbol{\cal{D}}} ({\bf{r}}) - {\boldsymbol{\cal{Z}}}
({\bf{r}}) - {\boldsymbol{\cal{I}}}_{m}  ({\bf{r}}).
\end{eqnarray}
The various fields in Eqs. (A4) - (A6) have the same nomenclature
and definitions in terms of their respective quantal sources as
given in the text.  There is, however, a fundamental difference
between the law for spin-less electrons Eq. (A4) and that for
electrons with spin Eq. (42).  This occurs in the Lorentz
${\boldsymbol{\cal{L}}} ({\bf{r}})$ and internal magnetic
${\boldsymbol{\cal{I}}}_{m}  ({\bf{r}})$ field components.  The
quantal source of these fields -- the physical current density
${\bf{j}} ({\bf{r}})$ -- is, in this case, a sum of the paramagnetic
${\bf{j}}_{p} ({\bf{r}})$ and diamagnetic ${\bf{j}}_{d} ({\bf{r}})$
components.

For electrons with spin, (the Schr{\"o}dinger-Pauli equation), one
could employ the same methodology as described above to arrive at
the corresponding `Quantal Newtonian' first law of Eq. (42). Instead
of providing this derivation, the law can be more easily derived via
comparison by writing the Hamiltonian $\hat{H}_\mathrm{spin-less}$
in terms of the current density ${\bf{j}} ({\bf{r}})$.  Thus,
\begin{equation}
\hat{H}_\mathrm{spin-less} = \hat{T} + \frac{1}{c} \int {\bf{j}}
({\bf{r}}) \cdot {\bf{A}} ({\bf{r}}) d {\bf{r}} - \frac{1}{2c^{2}}
\int \rho ({\bf{r}}) A^{2} ({\bf{r}}) d {\bf{r}} + \hat{W} +
\hat{V},
\end{equation}
with $\hat{T}$ the canonical kinetic energy operator and ${\bf{j}}
({\bf{r}}) = {\bf{j}}_{p} + {\bf{j}}_{d} ({\bf{r}})$. However, Eq.
(A7) is of the same form as the Hamiltonian $\hat{H}$ of the
Schr{\"o}dinger-Pauli equation Eq. (16).  The only difference
between these two equations is that in the latter, the current
density ${\bf{j}} ({\bf{r}}) = {\bf{j}}_{p} + {\bf{j}}_{d}
({\bf{r}})  + {\bf{j}}_{m} ({\bf{r}})$.  Thus, the resulting
`Quantal Newtonian' first law  Eq. (42) is also of the same form as
Eq. (A4) but with the added contribution of the magnetization
current density ${\bf{j}}_{m} ({\bf{r}})$ in the Lorentz
${\boldsymbol{\cal{L}}} ({\bf{r}})$ and internal magnetic
${\boldsymbol{\cal{I}}}_{m}  ({\bf{r}})$ field components.  Thus,
the `Quantal Newtonian' first law of Schr{\"o}dinger-Pauli theory is
derived.

The `Quantal Newtonian' first law of Eq. (65) for the model
noninteracting fermionic system possessing the same $\{ \rho
({\bf{r}}), {\bf{j}} ({\bf{r}}), N, {\bf{L}}, {\bf{S}} \}$ as that
of the interacting Schr{\"o}dinger-Pauli system can be derived by
writing the single-particle orbitals $\phi_{j} ({\bf{r}}) =
\phi_{j}^{R} ({\bf{r}}) + i \phi_{j}^{I} ({\bf{r}})$, where
$\phi^{R}_{j} ({\bf{r}})$ and $\phi^{I}_{j} ({\bf{r}})$ are the real
and imaginary parts, substituting in the differential equation Eq.
(64), and employing the continunity condition.

It can also be obtained by recognizing that the Schr{\"o}dinger
theory and Schr{\"o}dinger-Pauli theory Hamiltonians $\hat{H}_{s}$
of the model noninteracting fermionic system (See Eq. (60)) are of
the same form.  The difference between the two lies in the fact that
in addition to the paramagnetic ${\bf{j}}_{p} ({\bf{r}})$ and
diamagnetic ${\bf{j}}_{d} ({\bf{r}})$ components, there is the
presence of the magnetization current density ${\bf{j}}_{m, s}
({\bf{r}})$ in the physical current density ${\bf{j}} ({\bf{r}})$,
and thus in the Hamiltonian $\hat{H}_{s}$, of the latter.  Of course
the corresponding differential equations, their solutions $\phi_{j}
({\bf{r}})$ and the resulting fields of the two model systems
differ.  But the form of the `Quantal Newtonian' first law is the
same.

\end{document}